\documentclass[twocolumn]{aastex62}

\shorttitle{$\rm H_2CO$ and $\rm H110\alpha$ observations toward Aquila molecular cloud}
\shortauthors{Komesh et al.}

\usepackage{romannum}

\def\kms {km\,s$^{-1}$}
\def\Rn2 {\Romannum{2} }

\begin{document}

\title{$\rm H_2CO$ and $\rm H110\alpha$ Observations toward the Aquila Molecular Cloud}

\correspondingauthor{Toktarkhan Komesh}
\email{toktarhan@xao.ac.cn}

\author{Toktarkhan Komesh}
\affil{Xinjiang Astronomical Observatory, Chinese Academy of Sciences, Urumqi 830011, PR China}
\affil{Department of Solid State Physics and Nonlinear Physics, Faculty of Physics and Technology, Al-Farabi Kazakh National University, Almaty, 050040, Kazakhstan}
\affil{University of the Chinese Academy of Sciences, Beijing 100080, P. R. China}

\author{Jarken Esimbek}
\affiliation{Xinjiang Astronomical Observatory, Chinese Academy of Sciences, Urumqi 830011, PR China}
\affiliation{Key Laboratory of Radio Astronomy, Chinese Academy of Sciences, Urumqi 830011, PR China}

\author{Willem Baan}
\affiliation{Xinjiang Astronomical Observatory, Chinese Academy of Sciences, Urumqi 830011, PR China}
\affiliation{Netherlands Institute for Radio Astronomy,  ASTRON, 7991 PD, Dwingeloo, The Netherlands}

\author{Jianjun Zhou}
\affiliation{Xinjiang Astronomical Observatory, Chinese Academy of Sciences, Urumqi 830011, PR China}
\affiliation{Key Laboratory of Radio Astronomy, Chinese Academy of Sciences, Urumqi 830011, PR China}

\author{Dalei Li}
\affiliation{Xinjiang Astronomical Observatory, Chinese Academy of Sciences, Urumqi 830011, PR China}
\affiliation{Key Laboratory of Radio Astronomy, Chinese Academy of Sciences, Urumqi 830011, PR China}

\author{Gang Wu}
\affiliation{Xinjiang Astronomical Observatory, Chinese Academy of Sciences, Urumqi 830011, PR China}
\affiliation{Key Laboratory of Radio Astronomy, Chinese Academy of Sciences, Urumqi 830011, PR China}
\author{Yuxin He}
\affiliation{Xinjiang Astronomical Observatory, Chinese Academy of Sciences, Urumqi 830011, PR China}
\affiliation{Key Laboratory of Radio Astronomy, Chinese Academy of Sciences, Urumqi 830011, PR China}
\author{Serikbek Sailanbek}
\affil{Xinjiang Astronomical Observatory, Chinese Academy of Sciences, Urumqi 830011, PR China}
\affil{Department of Solid State Physics and Nonlinear Physics, Faculty of Physics and Technology, Al-Farabi Kazakh National University, Almaty, 050040, Kazakhstan}
\affil{University of the Chinese Academy of Sciences, Beijing 100080, P. R. China}
\author{Xindi Tang}
\affiliation{Xinjiang Astronomical Observatory, Chinese Academy of Sciences, Urumqi 830011, PR China}
\affiliation{Key Laboratory of Radio Astronomy, Chinese Academy of Sciences, Urumqi 830011, PR China}

\author{Arailym Manapbayeva}
\affil{Department of Solid State Physics and Nonlinear Physics, Faculty of Physics and Technology, Al-Farabi Kazakh National University, Almaty, 050040, Kazakhstan}

\begin{abstract}

The formaldehyde $\rm H_2CO(1_{10} - 1_{11})$ absorption line and H$110\alpha$ radio recombination line (RRL) have been observed 
toward the Aquila Molecular Cloud using the Nanshan 25\,m telescope operated by the Xinjiang Astronomical Observatory CAS. 
These first observations of the $\rm H_2CO$ $(1_{10} - 1_{11})$ absorption line determine the extent of the molecular regions that are affected by the ongoing star formation in the Aquila molecular complex and show some of the dynamic properties. 
The distribution of the excitation temperature $T_{ex}$ for $\rm H_2CO$ identifies the two known star formation regions W40 and Serpens South as well as a smaller new region Serpens 3. 
The intensity and velocity distributions of $\rm H_2CO$ and $\rm ^{13}CO(1-0)$  do not agree well with each other, which confirms that the $\rm H_2CO$ absorption structure is mostly determined by the excitation of the molecules resulting from the star formation rather than by the 
availability of molecular material as represented by the  distribution.
Some velocity-coherent linear $\rm ^{13}CO(1-0)$ structures have been identified in velocity channel maps of $\rm H_2CO$ and it is found that the three star formation regions lie on the intersect points of filaments.
The $\rm H110\alpha$ emission is found only at the location of the W40 H\,{\scriptsize II} region and spectral profile indicates a redshifted spherical outflow structure in the outskirts of the H\,{\scriptsize II} region.
Sensitive mapping of $\rm H_2CO$ absorption of the Aquila Complex has correctly identified the locations of star-formation activity in complex molecular clouds and the spectral profiles reveal the dominant velocity components and may identify the presence of outflows.

\end{abstract}


\section{Introduction}
The Aquila Molecular Cloud (AMC) or the Aquila Rift complex is located along the Galactic plane and stretches 
from 20\degr\, to 40\degr\, in longitude and from -1\degr\, to 10\degr\, in latitude, as revealed by CO and H\,\Romannum{1},
observations \citep{Dame2001,Prato2008P}. The western part of Aquila Rift contains several active star-forming 
regions: Serpens Main, Serpens South, W40, and MWC\,297. Here, we focus on part of the Aquila Rift complex that 
harbors two known sites of star formation: the western Serpens South is a young embedded 
cluster \citep{Bontemps2010}, and the eastern W40 is  a cluster associated with an H\,{\scriptsize II}  region 
\citep{Smith1985}. 

Spitzer observations show W40 and the embedded cluster Serpens South nearby on the sky such that Serpens South and is seemingly part of the W40 region \citep{Gutermuth2008}.
The distance to Serpens Main and W40 has recently been measured to be 436 pc and the distance to 
Serpens South should be similar, because the velocities of these clouds are very similar \citep{2017ApJ...834..143O}. 
Similarly \citet{2018PASJ..tmp..131S} proposed that Serpens South region may be interacting with the W40 HII expanding shell, which would place Serpens South and W40 at almost the same distance.
However, wide-field observations of the  $\rm ^{12}CO$ $(2-1)$ and  $\rm ^{13}CO$ $(2-1)$ emissions towards the Aquila Rift show 
the two spatially-extended components Serpens South and W40 at different velocities, which suggests that arcs and large-scale expanding bubbles and/or flows affect the velocity fields and play a role in the formation 
and evolution of the these clouds \citep{2017ApJ...837..154N}.

In this study we consider the $\rm H_2CO$ absorption at 4.830 GHz and $\rm H110\alpha$ recombination line emission at 
4.874 GHz in the Aquila Rift. 
The distribution of $\rm H_2CO$ absorption in the Galaxy and against 262 Galactic radio sources has been surveyed 
showing that $\rm H_2CO$ absorption is associated with most of the H\,{\scriptsize II} regions \citep{1979IAUS...84...81D, 
Downes1980, Pipenbrink1988}. 
Since $\rm H_2CO$ is only seen in absorption against a background continuum it only samples the physical conditions in the 
foreground of the H\,{\scriptsize II} region, while other mm and sub-mm spectral lines are observed both in front and behind
the source. The correlation between the distribution of $\rm ^{13}CO$ emission and $\rm H_2CO$  absorption is found to be very 
strong such that both components arise from similar regions \citep{2013A&A...551A..28T}.

In the present paper, we show results of the first $\rm H_2CO$ and $\rm H110\alpha$ observations toward the W40 and Serpens 
South regions of the Aquila Rift. In Section 2, we present the details of our observations. 
The results and discussion of the observations are described in Section 3. 
Finally, the conclusions are summarised in Section 4.

\section{Observations and database archives}

\subsection{$\rm H_2CO$ and $\rm H110\alpha$ observations}

The $\rm H_2CO$ $(1_{10}$-$1_{11})$  absorption line ($\lambda$ = 6 cm, $\nu_0=4829.6594$ MHz) and the 
$\rm H110\alpha$ RRL ($\nu_0=4874.1570$ MHz)  have been observed in the Aquila molecular cloud during February 2015 
using the Nanshan 25-m radio telescope of the Xinjiang Astronomical Observatory of Chinese Academy of Sciences. 
The 25-m radio telescope has an HPBW (half power beam width) of 10\arcmin\, at this wavelength. 
The observations were performed in an On-The-Fly mode with an average integration time of one minute for each position. 
The central position of the observing pattern is $18^h30^m03^s$ -2\degr02\arcmin40\arcsec (J2000). 
The 6 cm low noise receiver had a system temperature of about 23 K during the observations. 
In order to observe the $\rm H_2CO$ and the $\rm H110\alpha$ RR lines simultaneously, the center 
frequency of the spectrometer was set at 4851.9102 MHz. A Digital Filter Bank was used with 8192 channels 
and 64 MHz bandwidth,  corresponding to a velocity resolution of 0.48 km $s^{-1}$ at 4.852 GHz. 
The sensitivity of the system (DPFU, Degrees Per Flux Unit) was 0.116 K $Jy^{-1}$ and the main beam efficiency 
at this wavelength is 65\%.

We used CLASS and GREG (parts of GILDAS) to process the $\rm H_2CO$ and $\rm H110\alpha$ line data.  The area of 
the Aquila molecular cloud observed is 100\arcmin $\times$ 100\arcmin. 
The average sigma noise level of these maps is 0.020 K.
The signal to noise ratio of all detected points was better than 3. 
Assuming a distance of 436 pc for the Aquila complex, the spatial scale of the maps is 0.124 pc arcmin$^{-1}$.

\subsection{Archival data}
The $\rm ^{13}CO(1-0)$ and  $\rm ^{12}CO(1-0)$ data observed with the 13.7 m millimeter wave telescope of 
Purple Mountain Observatory in Delingha in April and May 2011 have been taken from the Millimeter Wave Radio 
Astronomy Database\footnote{http://www.radioast.nsdc.cn}. 
The velocity resolution of this data is 0.17 \kms and the system temperature of these on-the-fly mode observations 
ranged from 250 to 310 K. 
The $\rm ^{13}CO(J=1-0)$ data to $10\arcmin$ has been resampled onto the $\rm H_2CO$ observing grid.
The sigma noise levels of the $\rm ^{13}CO(1-0)$ and  $\rm ^{12}CO(1-0)$ data are 0.056 K and 0.122 K, respectively.
The 6 cm continuum data for the Aquila Rift region has been obtained from the Sino-German $\lambda$6 polarization 
survey of the Galactic Plane using the Urumqi 25 m telescope of the National Astronomical Observatories, CAS, were taken by 
\citet{2011A&A...527A..74S}. The central frequency of the data was 4.8 GHz, and the observing bandwidth was 600 MHz. 
The resolution of the data is 9.5\arcmin and the system temperature was about 22 K at the zenith.

\begin{figure*}
	\gridline{\fig{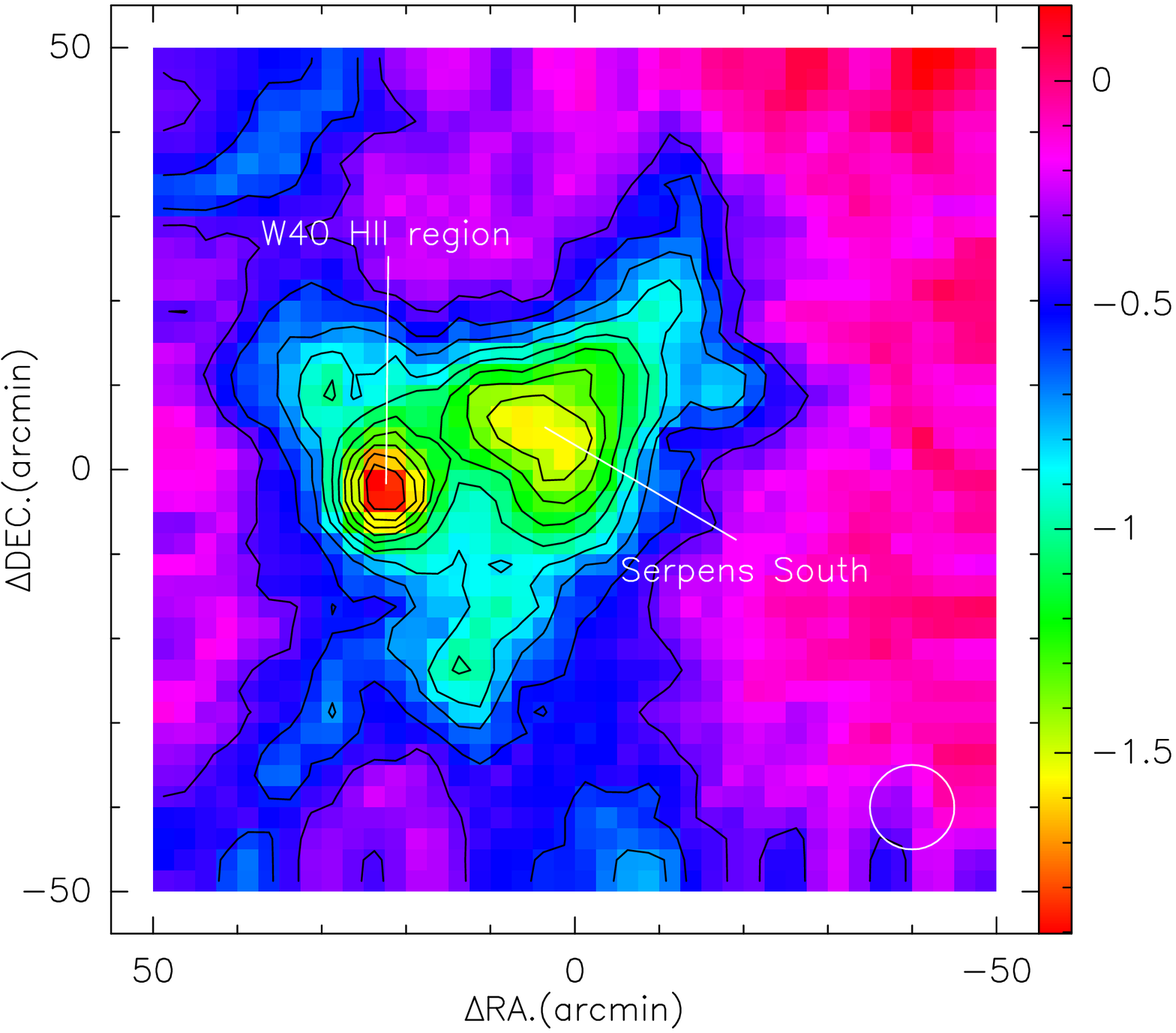}{0.47\textwidth}{(a)}
		\fig{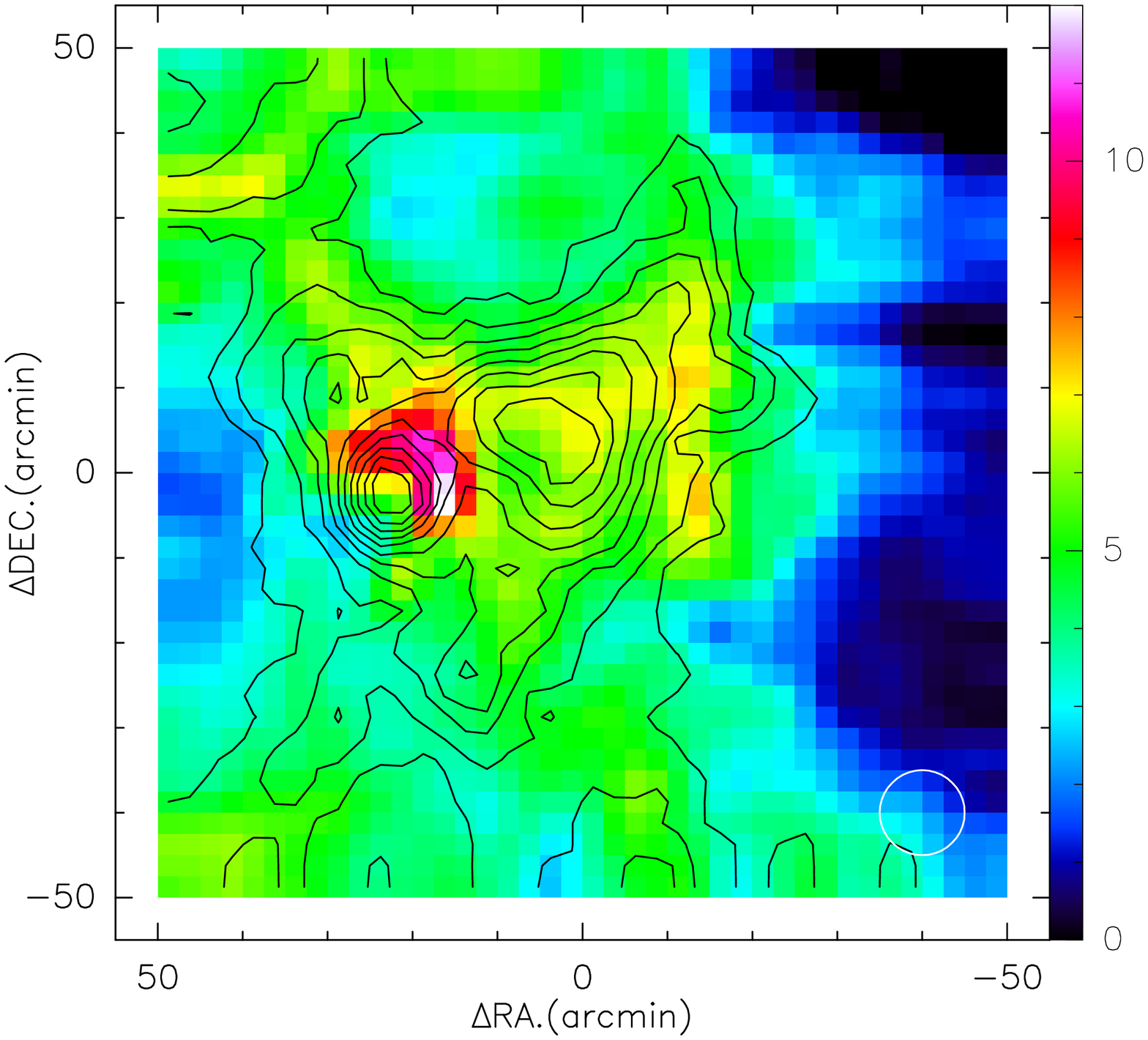}{0.47\textwidth}{(b)}
	}
	\caption{(a) A map of the integrated intensities of $\rm H_2CO$ absorption line and (b)  a map of the $\rm ^{13}CO(J=1-0)$ 
		integrated intensity superposed on contours of $\rm H_2CO$ toward Aquila molecular cloud. 
		The color bars are given in units of K km $s^{-1}$. Contour levels of the $\rm H_2CO$ intensity map are -0.4 to -1.8 in steps of 			-0.15 K km $s^{-1}$ for both panels.  White circles in the lower right illustrate the half-power beam size of 10\arcmin\,. 
		\label{fig:2}}
\end{figure*}

\begin{figure*}
	\gridline{\fig{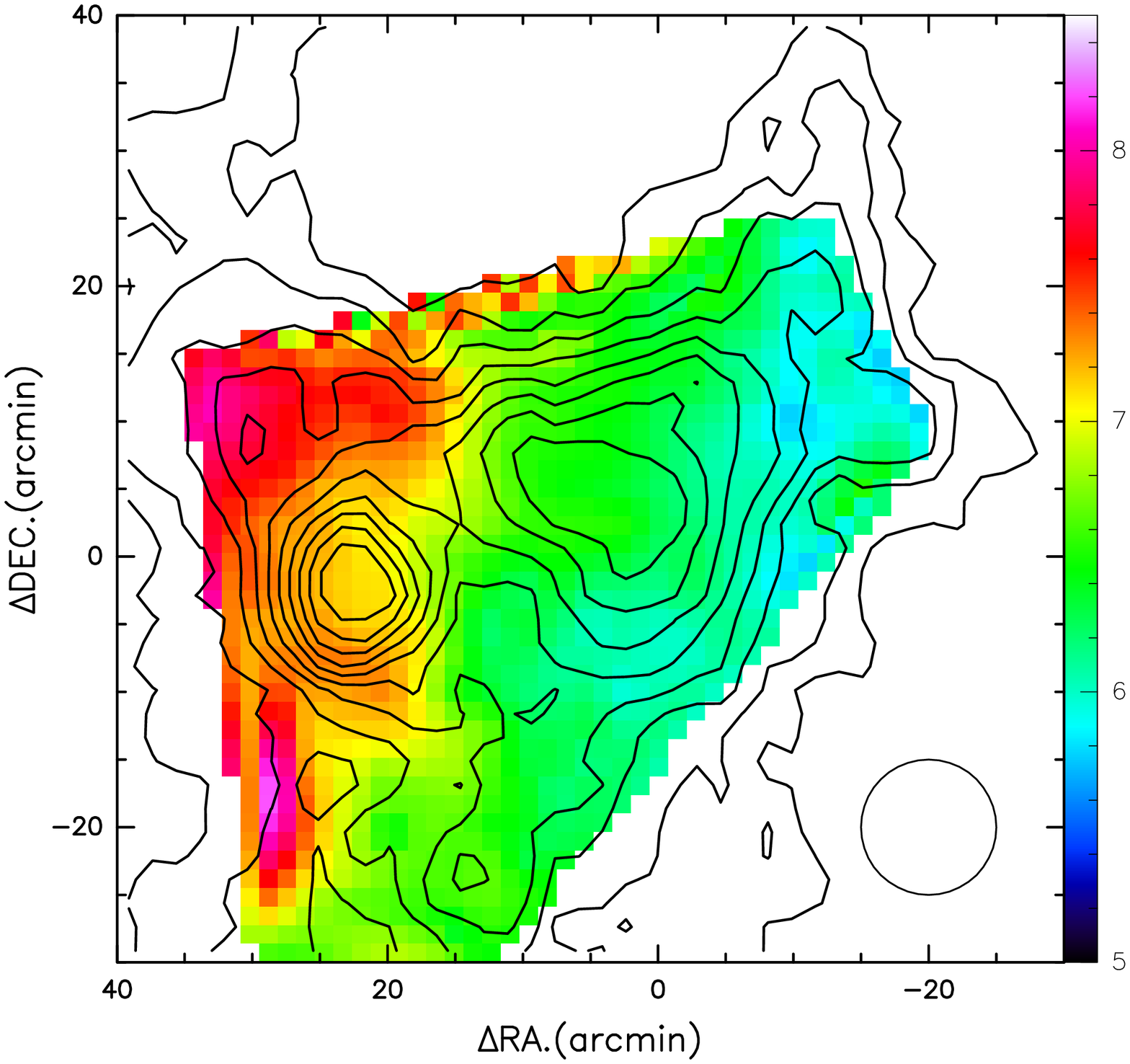}{0.47\textwidth}{(a)}\label{subfig:V_h2co}
		\fig{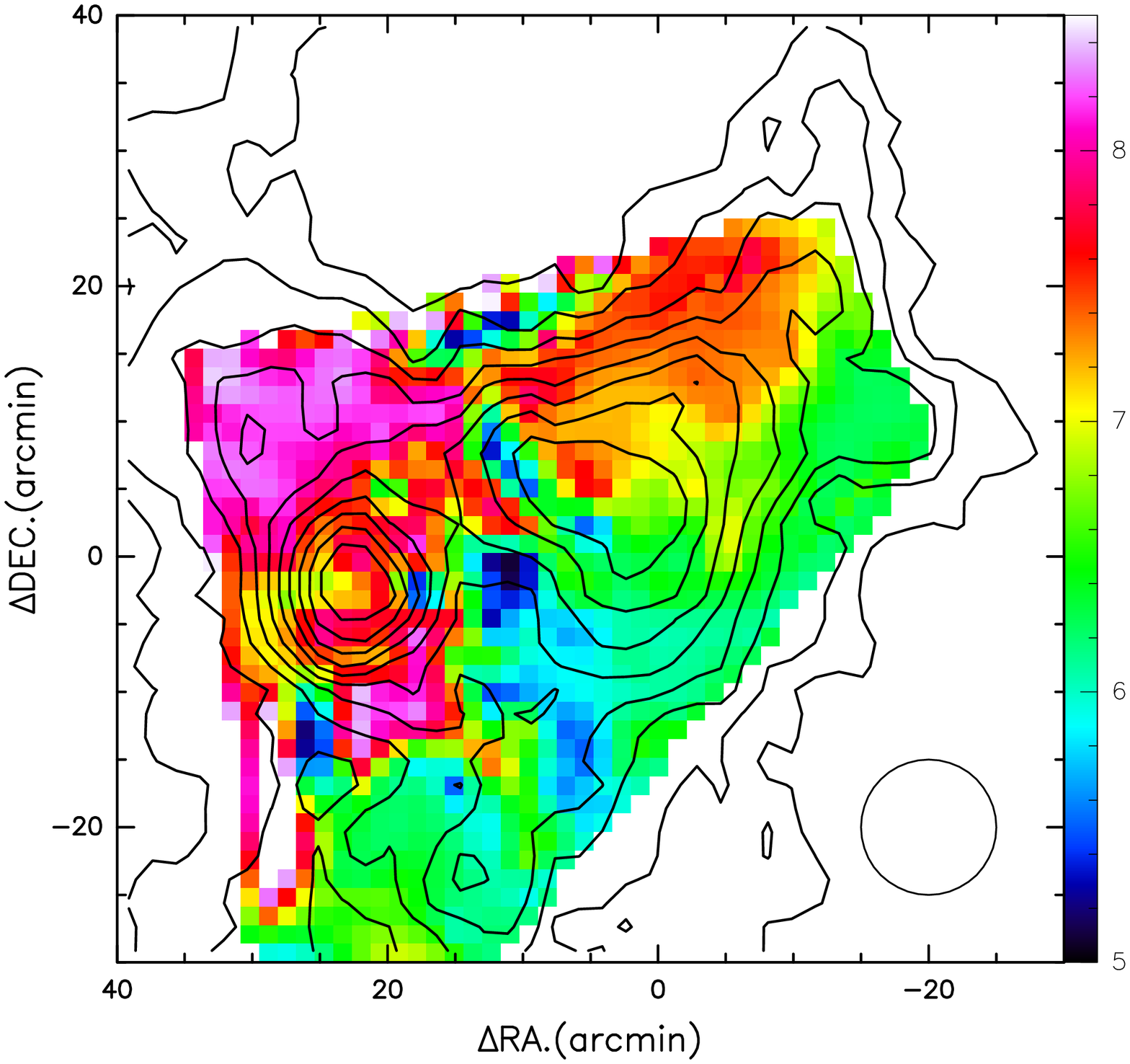}{0.47\textwidth}{(b)}\label{subfig:V_13co}
	}
	\caption{(a) The distribution of the centroid velocity of the $\rm H_2CO$ absorption
		and (b) of the $\rm ^{13}CO(1-0)$ emission superposed on the integrated intensity contours of $\rm H_2CO$. 
		The colour bars are given in units of \kms. The $\rm H_2CO$ contour levels are the same as those in Fig.1. Black circles in the lower right illustrate the half-power beam size of 10\arcmin\,. 
		\label{fig:dist_V}}
\end{figure*}

\begin{figure}
	\epsscale{1.2}
	\plotone{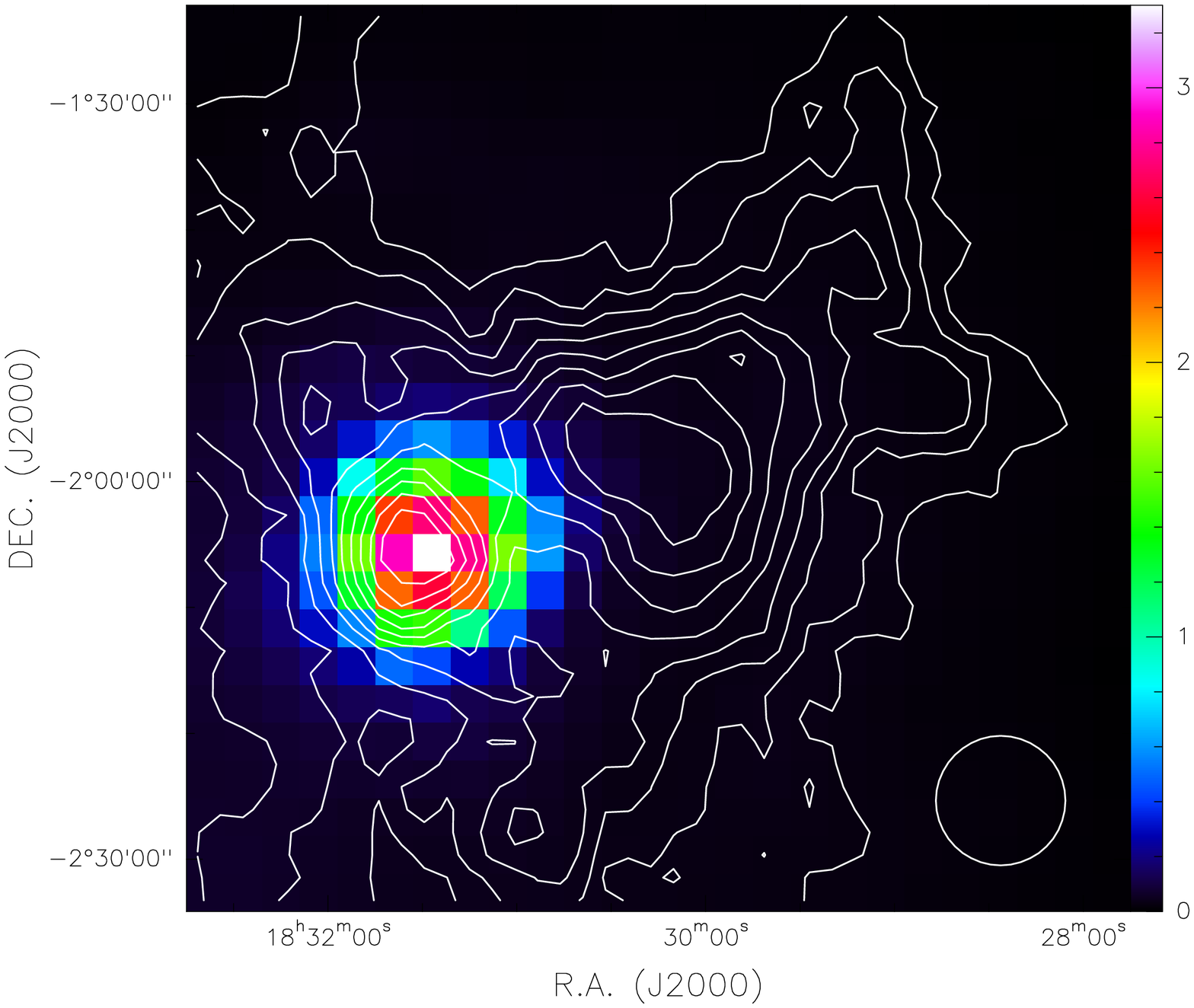}
	\caption{The 6 cm radio continuum distribution superposed on the $\rm H_2CO$ integrated absorption contours toward Aquila MC. 
		The $\rm H_2CO$ contour levels are the same as those in Fig.\ref{fig:2}. The temperature color bar is given in units of K. The white circle in the lower right illustrates the half-power beam size of 10\arcmin\,. 					\label{fig:Continuum}}
\end{figure}

\section{Results and discussion}

\subsection{The Formaldehyde Absorption}

The intensity map of the $\rm H_2CO$ absorption towards the Aquila Molecular Cloud integrated over the velocity interval 
of 3 - 11 km $s^{-1}$ is shown in Fig.\ref{fig:2}a. The observation parameters are shown in Tab.\ref{tbl:1}. 
Two concentrations can be seen in the map, that correspond to the W40 H\,{\scriptsize II} region and Serpens South 
with maximum flux values of the $\rm H_2CO$ absorption of -1.097 K \kms and -1.007 K km $s^{-1}$.   
The velocities at the offsets of (20,-5) and (0,0) are 7.125 \kms and 6.216 \kms, respectively.
The color-scale map of the $\rm ^{13}CO(J=1-0)$ emission is shown together with the $\rm H_2CO$ contours in Fig.\ref{fig:2}b. 
The $^{13}$CO emission map at Serpens South shows several elongated structures that only partially follow the $\rm H_2CO$ absorption structure and there is enhanced emission on the north 
side o W40 that could result from heating by the H\,{\scriptsize II} region.

The $\rm H_2CO$ absorption and the $\rm ^{13}CO$ emission structures are not in good agreement with each other in both the 
Serpens South and the H\,{\scriptsize II} region. The $^{13}$CO emission at W40 is offset and there is no clear concentration 
or a change in optical depth corresponding to Serpens South in Fig.\ref{fig:2}b. 
The velocity distributions of $\rm H_2CO$ is smooth and shows a gradient towards the East of the AMC (Fig.\ref{fig:dist_V}a). 
In the W40 region the velocity of $\thicksim7$ km/s corresponds to the systematic velocity of cold gas surrounding the entire region \citep{2015ApJ...806..201S} and agrees with the velocity found at the periphery of this region by \citet{2018PASJ..tmp..131S}. The velocity of $\rm H_2CO$ ( $\thicksim$ 6 km/s) at the southern part of main body of Serpens South agrees with the $^{13}$CO velocity distribution. 
On the other hand, the velocity structure of $\rm ^{13}CO$ (Fig.\ref{fig:dist_V}b) shows significant substructure and a gradient 
towards the North-East. The low intensity regions at the boundary between W40 and Serpens South may be related to 
an interaction between the regions as suggested by  \citet{2018PASJ..tmp..131S}.
In detail, the $\rm H_2CO$ absorption and $\rm ^{13}CO$ emission display different distributions and there 
is no clear evidence that the Serpens South region is superposed on the sky with the W40 region.

The 5 GHz radio continuum image within AMC is displayed in Fig.\ref{fig:Continuum}. The continuum temperature T$_c$ in 
the H\,{\scriptsize II} region ranges from 1 to 3.3 K, and is less than 0.04 K in Serpens South (in Tab.\ref{tbl:1}, col.6).
The strong $\rm H_2CO$ absorption at W40 is clearly related to the radio continuum of the H\,{\scriptsize II} region, while 
the continuum is significantly weaker at Serpens South. Because the $^{13}$CO emission does not show any enhancement 
at Serpens South, the $\rm H_2CO$ absorption there should result from an enhanced continuum background within Serpens 
South in addition to the cosmic microwave background (CMB).
Therefore the absorption contours in W40 and Serpens South define the extent of the region that is affected by the ongoing 
star formation, which is well beyond the sites of star formation.

The formaldehyde line is in absorption across the whole region of W40 and Serpens South. This can only happen when 
everywhere the excitation temperature of the 6-cm line is less than the brightness temperature of the radio continuum sources 
plus the microwave background.
While the excitation temperature $T_{ex}$ may be uniformly low ($\leq$ 1K) in cold clouds \citep{1973ApJ...183..441H}, the 
excitation conditions may vary strongly across star formation regions. 
Therefore,  a simple determination of the absorbing $\rm H_2CO$ column density would not be accurate. Instead we perform an approximate determination of the excitation temperature of $\rm H_2CO$ across the region.
Making the imperfect assumption that the column densities of $\rm H_2CO$ and $\rm ^{13}CO(1-0)$ are correlated, we may estimate the $\rm H_2CO$ column density from the column density of $\rm ^{13}CO(1-0)$, which may be obtained from \citet{1994ApJ...435..279S}. 
Assuming an $\rm H_2CO$ to $H_2$ abundance ratio of $3\times10^{-9}$ \citep{1975ApJ...196..433E} and a $\rm ^{13}CO(1-0)$ 
to $H_2$ abundance ratio of $2.4\times10^{-6}$ \citep{2013MNRAS.431.1296R},  the column density ratio of  
between $\rm H_2CO$ and $\rm ^{13}CO(1-0)$ may be calculated as:
\begin{equation}
N(H_2CO)=1.25\times10^{-3}N(^{13}CO).
\end{equation}
Using the optical depth $\tau_{app}$ of $\rm H_2CO$ as calculated using the formulation in \citet{Pipenbrink1988}:
\begin{equation}
N(H_2CO) = 9.4 \times 10^{13} \cdot \tau_{app} \cdot \Delta V (cm^{-2}),
\end{equation}
which already assumed a mean value for $T_{ex}$ = 2 K.  The excitation temperature of $\rm H_2CO$ may then be calculated using \citet{1973ApJ...183..441H} as follows:
\begin{equation}
T_L=(T_{ex}-T_c)[1-exp(-\tau_{app})] (K),
\end{equation}
where $T_L$ is the antenna temperature of the absorption line in degrees kelvin, and $T_c$ is the continuum brightness 
temperature of the radio background plus the CMB. The results for $N(H_2CO)$ and $T_{ex}$ are presented in Tab.\ref{tbl:1}.

\begin{figure}
	\epsscale{1.2}
	\plotone{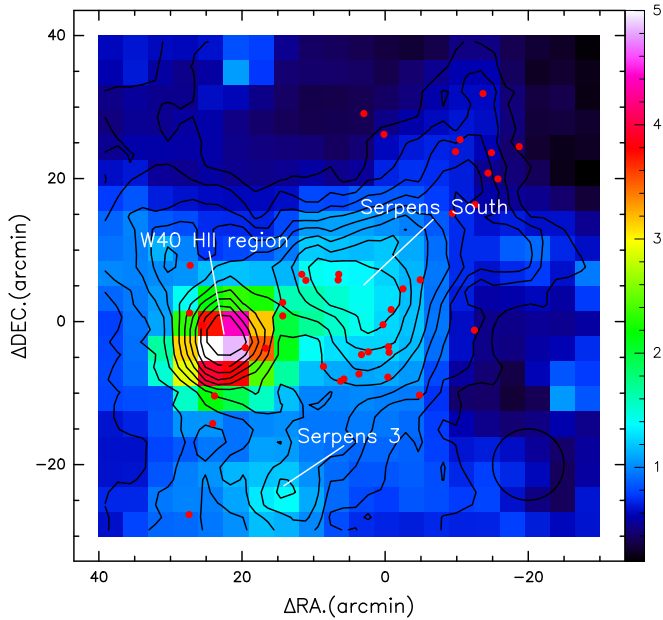}
	\caption{Distribution of $T_{ex}$ of $\rm H_2CO$ absorption line superposed on the integrated absorption contours of $\rm H_2CO$. The contour levels are the same as Fig.\ref{fig:2}. The $T_{ex}$ color bar is given in units of K. The black circle in the lower right illustrates the half-power beam size of 10\arcmin\,.  \label{fig:Tex}}
\end{figure}

The distribution of $T_{ex}$ determined in this manner shows an enhanced temperature around W40 H\,{\scriptsize II} region 
ranging from 2 - 5 K and around Serpens South ranging from 1 - 2 K (Fig.\ref{fig:Tex}). 
This enhanced $T_{ex}$ in Serpens South confirms the presence of local heating source and ongoing star formation activity. 
In addition, there is another Serpens 3 region in the formaldehyde absorption structure south of W40 that shows similar 
conditions of (prestellar) star-formation and inferred weak radio continuum (Fig.\ref{fig:Tex}).
It appears that the extent of the $\rm H_2CO$ absorption is determined mostly by the excitation of the molecules in regions that are affected by the star formation rather than by the availability of molecular material as represented by the $\rm ^{13}CO(J=1-0)$ distribution.
In order to connect the Serpens 3 region with evidence of ongoing star formation, the locations of protostellar cores \citep{2015A&A...584A..91K} in the region have been added to Fig. \ref{fig:Tex}. 
While there are protostellar cores associated with W40 and Serpens South, there are no cores (yet) present around Serpens 3. 
This would suggest that the Serpens 3 region is still less evolved than the other regions of star-formation. 
On the other hand, protostellar cores are found around the north-west extension of the $\rm H_2CO$ absorption, which shows less enhancement of $T_{ex}$ but may also qualify as a star-forming region.

\begin{figure*}
	\gridline{\fig{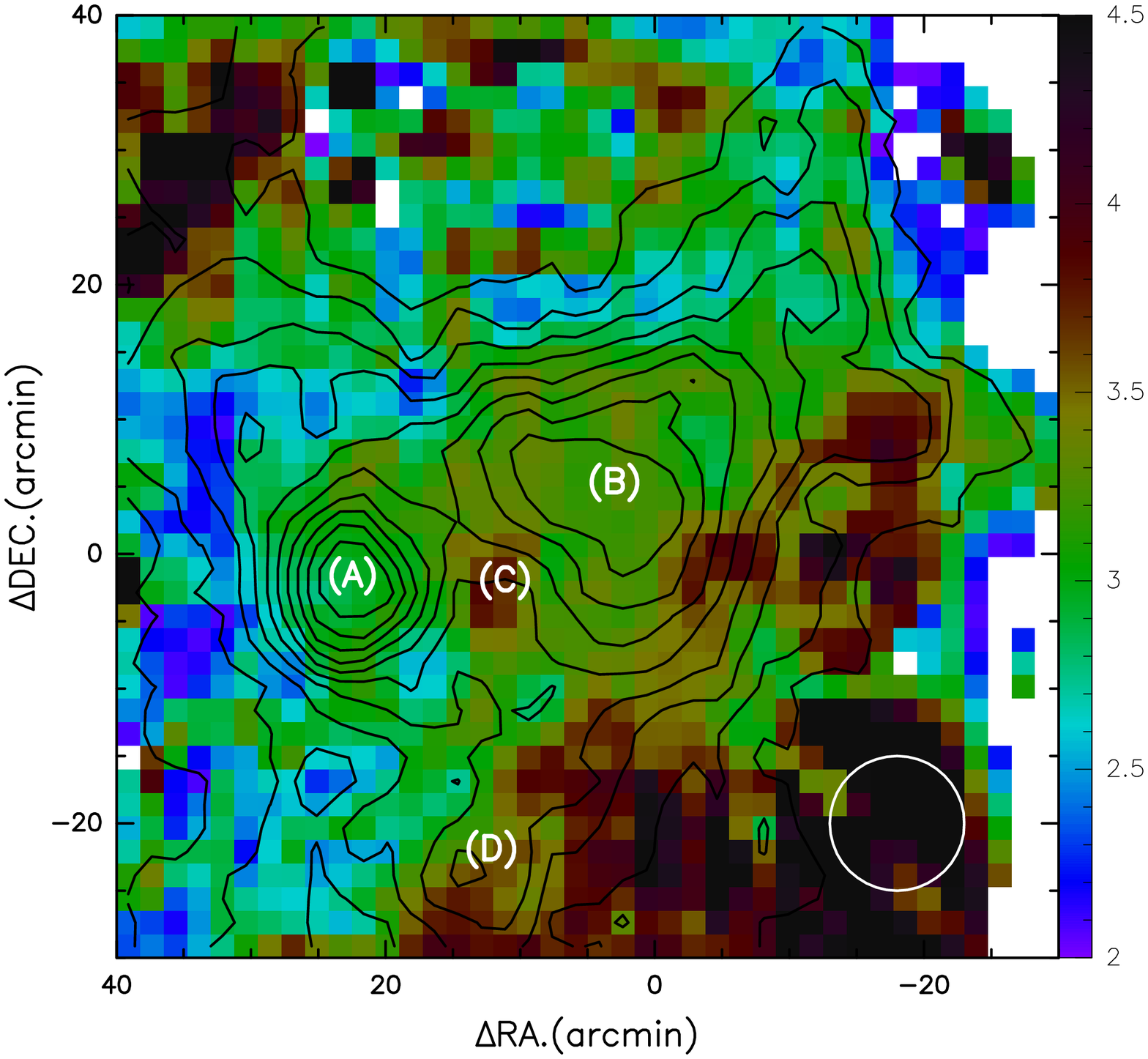}{0.47\textwidth}{}\label{subfig:V_h22co}
		\fig{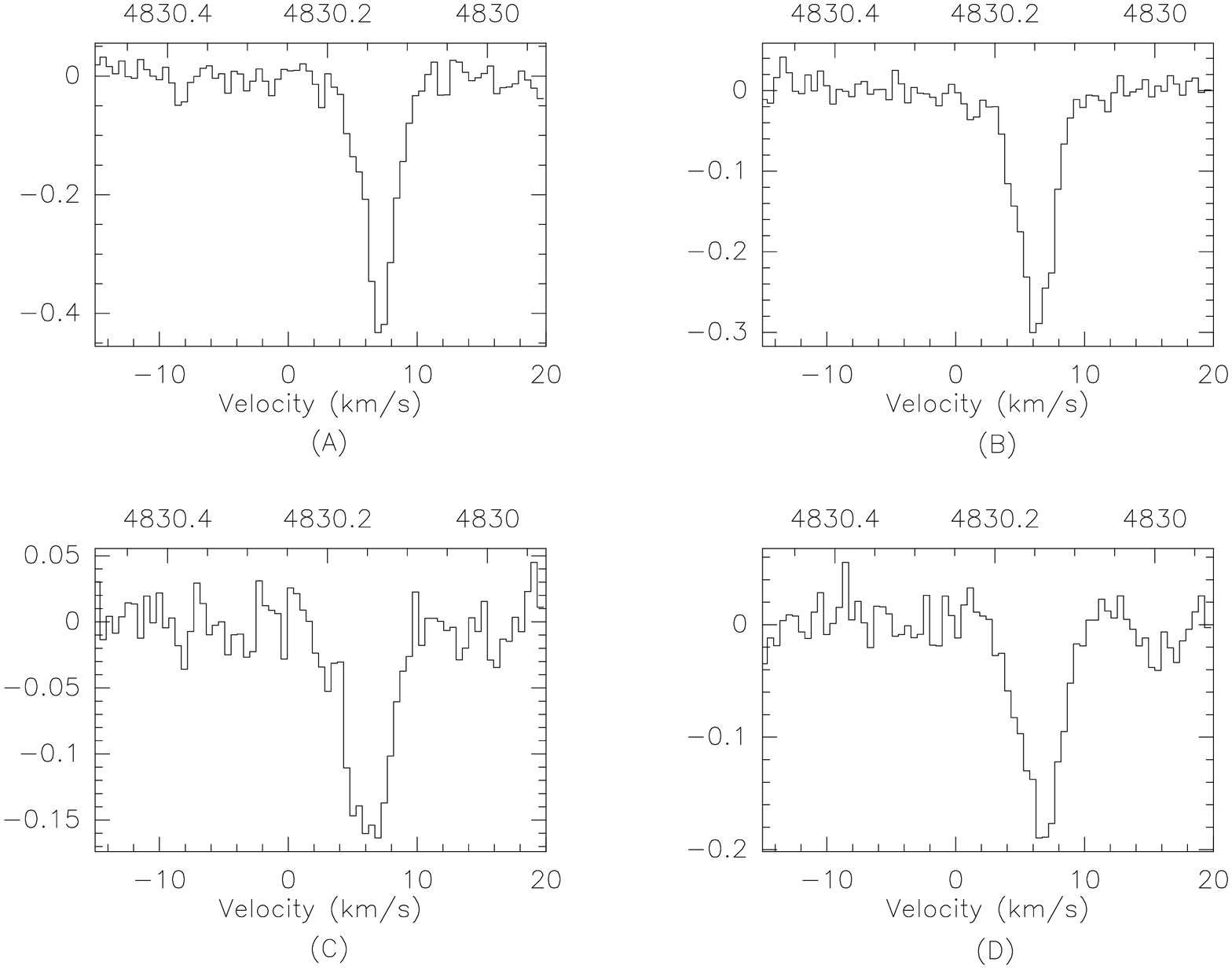}{0.5\textwidth}{}\label{subfig:V_132co}
	}
	\caption{The $\rm H_2CO$ line width map superposed on the $\rm H_2CO$ integrated intensity contours (left panel) and spectra 
	at four selected regions (right panels). The contour levels are the same as Fig.\ref{fig:2}. The line width color bar is given 		in units of \kms. The white circle in the lower right illustrates the half-power beam size of 10\arcmin\,. 				
	\label{fig:W_h2co}}
\end{figure*}

The map of the $\rm H_2CO$ line width superposed on the $\rm H_2CO$ integrated intensity contours and spectral plots at four 
selected regions are shown in Fig.\ref{fig:W_h2co}. The spectra at locations (A) around W40 H\,{\scriptsize II} and (B) 
around Serpens South show a single feature with a multicomponent substructure. 
The spectral feature at Serpens South is slightly weaker and broader than the feature at W40 and shows shallow 
redshifted and blueshifted wings, possibly resulting from outflows or the superposition of multiple components. 
The W40 spectrum shows a possible blueshifted feature in  at -8.5 \kms that could have a counterpart in the 
$\rm H110\alpha$ RL at that location (see section below).
The  spectra at locations (C) and (D) show a broader multicomponent structure. 
In the (C) region at a distance of of 0.7 pc West of the W40 H\,{\scriptsize II} the spectrum is broadened relative to the 
W40 spectrum (A) and has a low velocity shoulder similar to that found in the $\rm ^{13}CO(1-0)$ spectrum and the 
map of Figure \ref{fig:dist_V}b. This lower velocity component may suggest the presence of an outflow component 
associated with the W40 H\,{\scriptsize II} region. 
The southeastern region (D) of Serpens 3 also shows line broadening toward lower velocities as compared with
the spectrum at W40 (A) as well as a high-velocity wing that may result from a superposition of cloud 
filaments moving to the northeast. In addition, there may be a separate high-velocity component at 15 \kms.

\begin{figure*}
	\epsscale{1.1}
	\plotone{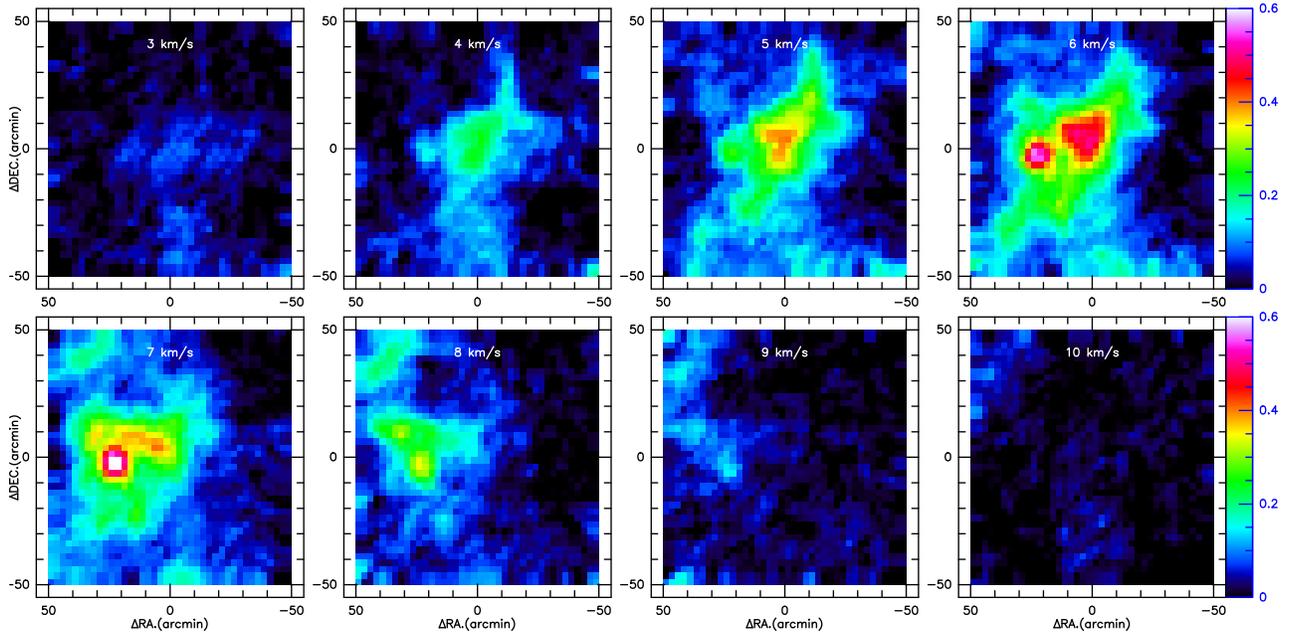}
	\caption{$\rm H_2CO$ velocity channel maps toward the Aquila Molecular Cloud. The color bar is in units 
	of K \kms.
	\label{fig:channel}}
\end{figure*}

\begin{figure}
	\epsscale{1.2}
	\plotone{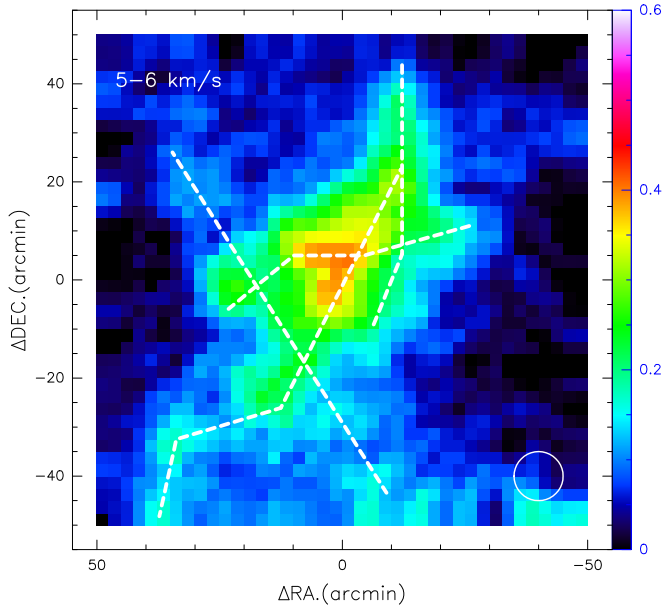}
	\caption{Intensity map of the $\rm H_2CO$ absorption integrated over the velocities of 5 - 6 km $s^{-1}$.
	The dashed lines represent filamentary structures  that may be discerned in the Aquila Molecular Cloud. The white circle in the lower right illustrates the half-power beam size of 10\arcmin\,. 				
	\label{fig:5-6}}
\end{figure}

The $\rm H_2CO$ channel maps are presented in Fig.\ref{fig:channel} with velocity interval of 1 \kms.  
Most of the Serpens South part has a velocity of 6 \kms, while most of the W40 part has velocity of 7 \kms. 
In the 5 \kms panel an east--west and a southeast--northwest velocity-coherent linear structure are present, which resemble  
the linear structures in $\rm ^{13}CO(1-0)$ intensity map (Fig.\ref{fig:2}b). Similarly at 8 \kms there is a northeast--southwest 
structure running through the W40 region.
In order to emphasize these structures, we indicate with dashed curves the locations of apparently velocity-coherent 
structures superposed on the $\rm H_2CO$ intensity map integrated over the intervals of 5 - 6 \kms in Fig.\ref{fig:5-6}. 
Assuming the distance of 436 pc, these linear structures are about 5 - 10 pc in length and they may be remnants of the 
super-bubbles converging in our observed area (see also \citet{2017ApJ...837..154N}).
It should be noted that our three star formation regions coincide with intersection points of these linear structures.

\begin{figure*}
\gridline{\fig{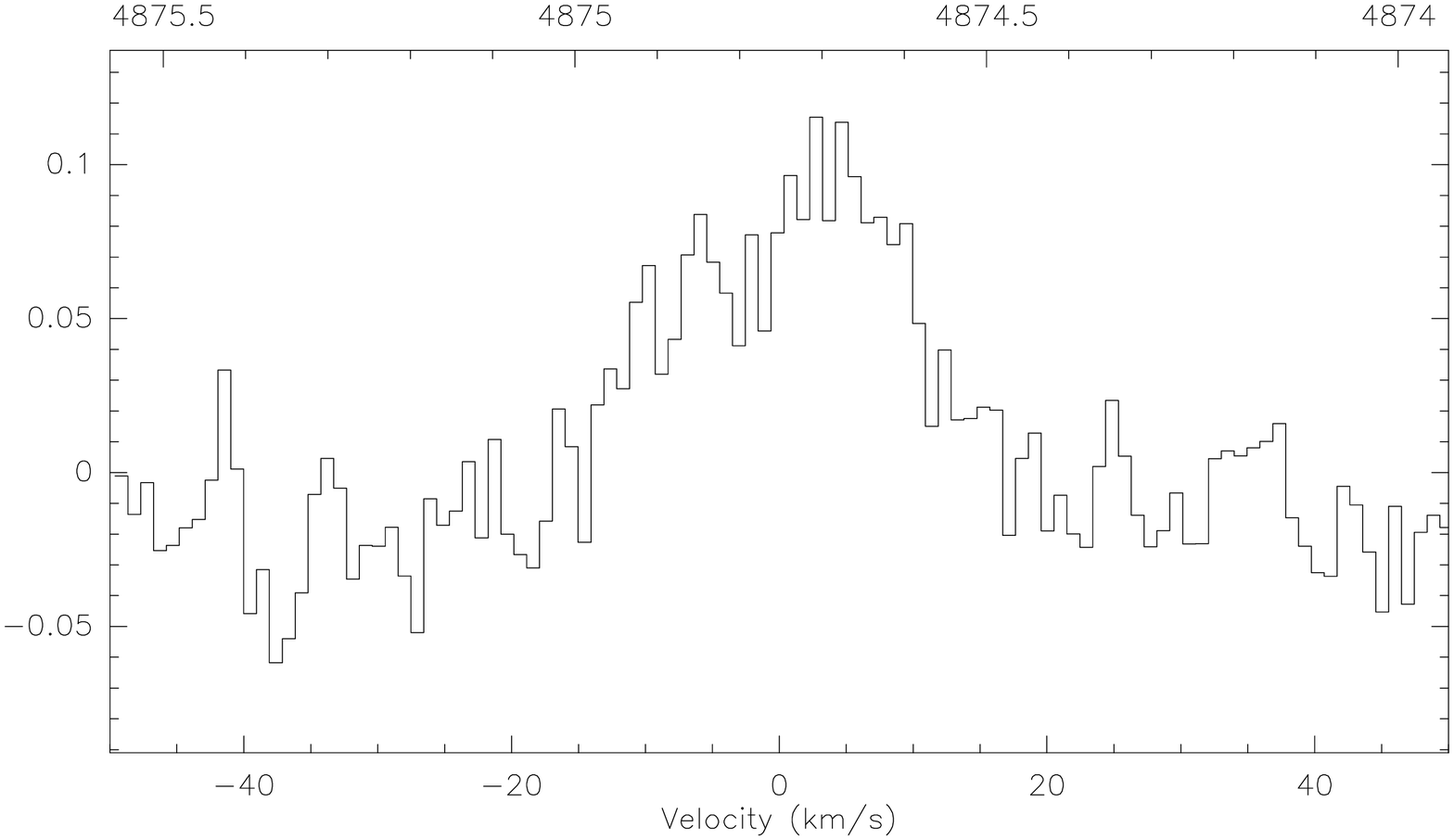}{0.47\textwidth}{(a)}
		\fig{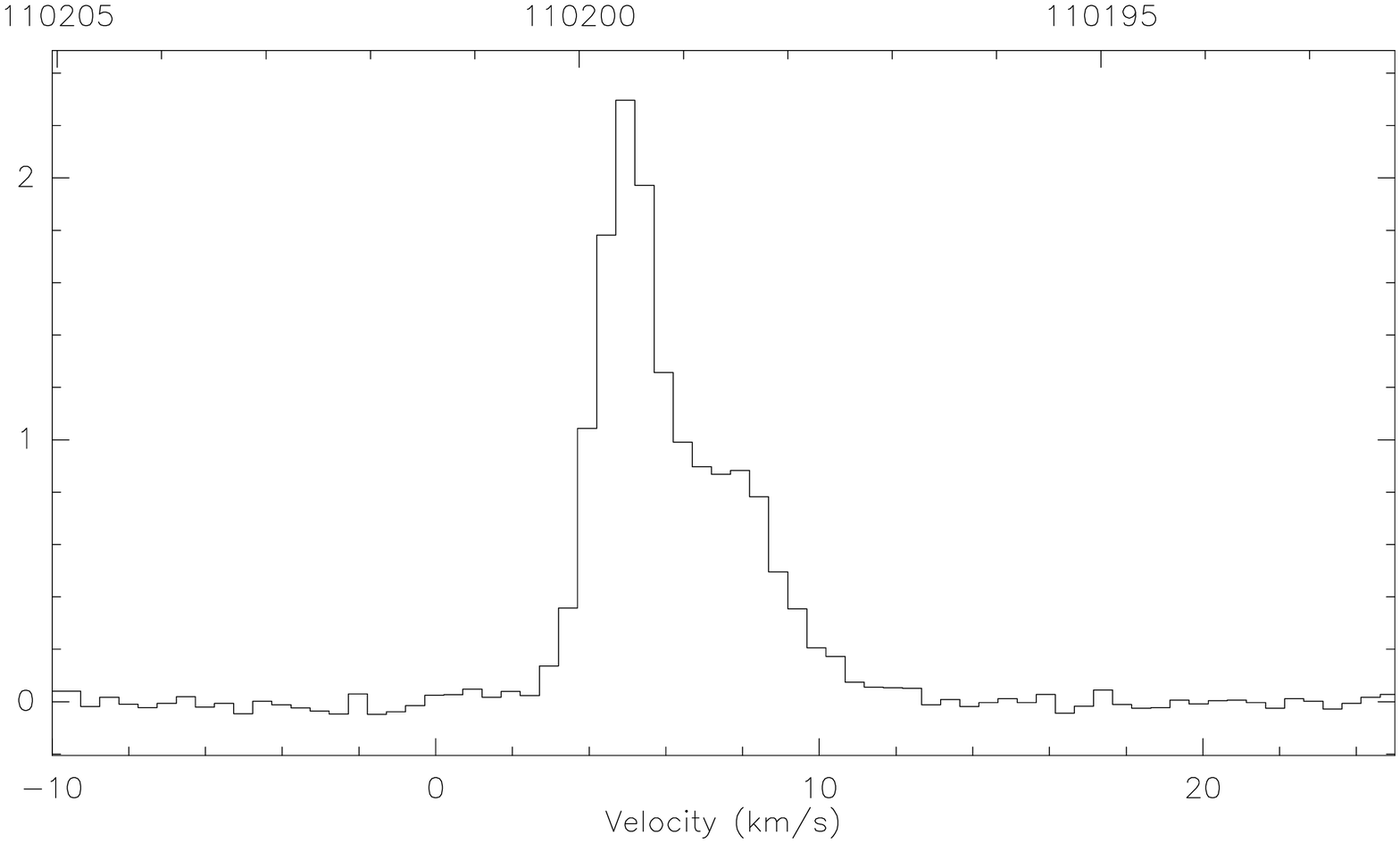}{0.47\textwidth}{(b)}
	}
	\caption{The spectra of (a) the $\rm H110a$ emission and (b) the  $\rm ^{13}CO(J=1-0)$ emission at the center of W40 H\,{\scriptsize II} 	region (region A in Fig.\ref{fig:W_h2co}).
	\label{fig:spec}}
\end{figure*}

\subsection{The $\rm H110\alpha$ Recombination Line}
  
The $\rm H110\alpha$ RRL is only detected at the W40 H\,{\scriptsize II} region. 
The spectrum of the $\rm H110\alpha$ line at W40 is presented in Fig.\ref{fig:spec} together with the $\rm ^{13}CO(1-0)$ profile 
at the same location. 
This much broader profile comes from the distribution across the circular image of W40, starting at 10 \kms and tapering off at -20 \kms. However, the $\rm H110\alpha$ spectrum at W40 does not have a counterpart in the $\rm ^{13}CO(1-0)$ profile at lower velocities, but there may be weak spectral component (to be confirmed) in the $\rm H_2CO$ absorption spectrum at -8.5 \kms.
With a peak continuum temperature of 3.33 K, the optical depth $(T_L/T_c)$ has a peak value of 0.038, with an average value of 0.15.
The profile of the $\rm H110\alpha$ line resembles a spherical (blueshifted) outflow with an approximate $(1- (V/V_o)^2)^n$ shape, where outflow velocity V$_o$ is on the order of 25 \kms.
The local thermodynamic equilibrium electron temperature $T^{*}_{e}$ is determined using recombination lines by \citet{1978ARA&A..16..445B} discussions as follows:
\begin{equation}
T^{*}_{e}=[3.624\times10^4 \cdot (T_c/(T_L \cdot \Delta v)]^{0.87} ,
\end{equation}
where the $T^{*}_{e}$ indicates a value of 7300 K assuming the peak optical depth and a total line width of 35 \kms.
This value is meaningful when all $\rm H110\alpha$ emission originates within the outskirts of a typical H\,{\scriptsize II} region.

\section{Conclusion}
The $\rm H_2CO$ ($1_{10} - 1_{11}$) absorption line and $\rm H110\alpha$ RRL emission line have been mapped for the 
first time toward the Aquila molecular cloud containing the W40 H\,{\scriptsize II} region and the Serpens South region. 
The map of the integrated intensity of the $\rm H_2CO$ absorption clearly defines the boundaries of the area that is affected 
by the embedded star formation. There is prominent absorption at the location of the W40 H\,{\scriptsize II} region and 
weaker absorption in the Serpens South region. 
A third star formation region Serpens 3 has been identified 1.4 pc south of W40. 
No substantial correspondence has been found between the three star formation regions and the emission map of $\rm ^{13}CO(1-0)$. 
Except for the W40 region, the intensity and velocity distributions of $\rm H_2CO$ and $\rm ^{13}CO(1-0)$ do not agree well 
with each other. 

The radio continuum structure of the region shows a $T_c$ peak at the W40 H\,{\scriptsize II} region, but only estimated values 
of 0.04 K in the Serpens South region and at Serpens 3. 
Therefore, the $\rm H_2CO$ absorption at W40 comes from the H\,{\scriptsize II} region plus the CMB background, while the weaker 
components at Serpens South and the Serpens 3 regions result from a weak continuum background plus the CMB.
The extent of the $\rm H_2CO$ absorption confirms that it is mostly determined by the excitation of the molecules in regions that are affected by the star formation rather than the availability of molecular gas as depicted by $\rm ^{13}CO(J=1-0)$.

Instead of determining the absorbing column density of the formaldehyde across the region with clearly varying 
environmental conditions, we determined the excitation temperature of the $\rm H_2CO$.
Assuming that the column densities of $\rm H_2CO$ relates to that of $\rm ^{13}CO(1-0)$ (see \citet{2013A&A...551A..28T}), 
and that there is fixed abundance ratio of $\rm H_2CO$ and $\rm ^{13}CO(1-0)$, the column density of $\rm ^{13}CO(1-0)$ may be 
used to determine the $T_{ex}$ distribution. While this procedure may not be very reliable, the $T_{ex}$ shows enhancements 
in the W40 H\,{\scriptsize II} region ranging from  2 to 5 K, and $>2$ at Serpens South. This also identifies the new star formation region 
Serpens 3 region with $T_{ex}$ = 2 K. 
The results show that the local conditions strongly affect the excitation of $\rm H_2CO$ and that assuming a constant value 
for $T_{ex}$ would not be appropriate across this region. 

The velocity structure of $\rm H_2CO$ is very smooth while the velocity structure of  $\rm ^{13}CO(1-0)$ has much fine structure 
and a gradient in a different direction.
Therefore, the presence of $\rm H_2CO$ absorption may be correlated with significant substructure within the $\rm ^{13}CO(1-0)$ integrated emission structure but there is no global correlation.
The integrated intensity map of $\rm ^{13}CO(1-0)$ shows several fine structure regions with discrepant velocities. 
One of these regions at about 1.2 pc west of W40 has a larger linewidth for $\rm H_2CO$ and shows a lower $\rm ^{13}CO(1-0)$ 
velocity. This region may be associated with an outflow found in a near-infrared survey by Zhang et al. (2015).

Some velocity-coherent filamentary structures have been identified in velocity channel maps of $\rm H_2CO$ that are possible 
remnants of earlier super-bubble structures. 
Assuming the distance of 436 pc, these linear structures range between about 5 - 10 pc in length. 
The three star formation regions are found to lie close to intersection points of these filaments, which may suggest a causal relation.
The observed velocity substructure of the $\rm H_2CO$ absorption lines may also relate to the presence of such filaments.

The $\rm H110\alpha$ RRL is only detected in the W40 H\,{\scriptsize II} region. The emission spectrum of $\rm H110\alpha$ shows a broad profile covering a velocity range of 30 \kms. The shape of the spectral profile suggests that this emission originates in a  shell
in front of W40 that is spherically expanding at $~$26 \kms. The LTE electron temperature corresponding to the line strength 
is estimated at 7900 K, which is typical value for the outskirts of an H\,{\scriptsize II} region. 

Sensitive mapping of $\rm H_2CO$ absorption has been able to correctly identify star-formation activity in complex molecular clouds 
such as the Aquila Complex. In addition, the detailed structure of the absorption lines may reveal discrepant velocity components resulting from outflow regions.

\acknowledgments
This work was sponsored by CAS-TWAS President's Fellowship for International Doctoral Students, and The National Natural Science foundation of China under grants 11433008, 11703074, 11703073, and 11603063, and The Program of the Light in China's Western Region (LCRW) under grant Nos. 2016-QNXZ-B-22, 2016-QNXZ-B-23, and 2018-XBQNXZ-B-024. W.A.B. has been funded by Chinese Academy of Sciences President's International Fellowship 
Initiative under grant No. 2019VMA0040.
This work is based on observations made with the Nanshan 25m radio telescope, which is operated by the Kay Laboratory of Radio Astronomy, Chinese Academy of Sciences.

\startlongtable
\begin{deluxetable}{r|rrrrrrrr}
	\tablecaption{Parameters of the  $\rm H_2CO$  $(1_{10} - 1_{11})$ absorption line\label{tbl:1}}
	\tablehead{\colhead{Offset}                    &
		\colhead{Flux}                    &
		\colhead{Velocity}                  &
		\colhead{Width}        &
		\colhead{$T_L$}                 &
		\colhead{$T_C$ }                &
		\colhead{$\tau_{app}$}                &
		\colhead{$N(H_2CO)$}                 &
		\colhead{$T_{ex}$}  \\
		\colhead{arcmin}  &
		\colhead{K km $s^{-1}$} &
		\colhead{km $s^{-1}$}		&
		\colhead{km $s^{-1}$}		&
		\colhead{K}	&
		\colhead{K}	&
		\colhead{}&
		\colhead{$10^{12}$$cm^{-2}$}	&
		\colhead{K} 
	}
	\colnumbers
	\startdata
	-25,-30   &		-0.2376  (0.02)				&		7.187			(0.18)	&			4.517		(0.52)	&			-0.049			&			0.011 			&			0.08      				&			3.19    			&		0.69      				\\
	-15,-30   &		-0.2935  (0.02)				&		5.66 			(0.24)	&			5.95 		(0.49)	&			-0.046			&			0.024 			&			0.08      				&			4.20    			&		0.67      				\\
	-10,-30   &		-0.2947  (0.02)				&		5.898			(0.15)	&			4.279		(0.33)	&			-0.065			&			0.027 			&			0.11      				&			4.40    			&		0.65      				\\
	-5,-30   &		-0.3709  (0.02)				&		5.647			(0.10)	&			3.853		(0.21)	&			-0.090			&			0.029 			&			0.17      				&			6.06    			&		0.62      				\\
	0,-30   &		-0.3654  (0.02)				&		5.883			(0.14)	&			4.114		(0.28)	&			-0.083			&			0.031 			&			0.14      				&			5.54    			&		0.66      				\\
	5,-30   &		-0.3531  (0.02)				&		6.193			(0.12)	&			3.835		(0.26)	&			-0.087			&			0.036 			&			0.15      				&			5.24    			&		0.68      				\\
	10,-30   &		-0.45    (0.02)				&		6.418			(0.11)	&			4.106		(0.27)	&			-0.103			&			0.043 			&			0.12      				&			4.77    			&		0.93      				\\
	15,-30   &		-0.5038  (0.03)				&		6.709			(0.10)	&			4.065		(0.23)	&			-0.116			&			0.046 			&			0.12      				&			4.48    			&		1.09      				\\
	20,-30   &		-0.4239  (0.03)				&		6.345			(0.12)	&			3.844		(0.27)	&			-0.104			&			0.048 			&			0.12      				&			4.16    			&		1         				\\
	25,-30   &		-0.3951  (0.03)				&		6.704			(0.13)	&			3.851		(0.30)	&			-0.096			&			0.051 			&			0.12      				&			4.33    			&		0.91      				\\
	30,-30   &		-0.4358  (0.01)				&		6.626			(0.05)	&			3.274		(0.11)	&			-0.125			&			0.061 			&			0.17      				&			5.20    			&		0.87      				\\
	35,-30   &		-0.3491  (0.02)				&		6.514			(0.07)	&			2.434		(0.14)	&			-0.135			&			0.072 			&			0.2       				&			4.56    			&		0.82      				\\
	40,-30   &		-0.2275  (0.02)				&		6.627			(0.10)	&			2.412		(0.24)	&			-0.089			&			0.075 			&			0.19      				&			4.24    			&		0.59      				\\
	-15,-25   &		-0.3384  (0.03)				&		5.586			(0.26)	&			6.38 		(0.68)	&			-0.050			&			0.024 			&			0.07      				&			4.01    			&		0.79      				\\
	-10,-25   &		-0.3575  (0.03)				&		6.19 			(0.16)	&			4.482		(0.36)	&			-0.075			&			0.029 			&			0.11      				&			4.58    			&		0.76      				\\
	-5,-25   &		-0.4409  (0.02)				&		5.608			(0.12)	&			4.842		(0.33)	&			-0.086			&			0.032 			&			0.11      				&			5.12    			&		0.84      				\\
	0,-25   &		-0.3756  (0.02)				&		5.548			(0.12)	&			4.069		(0.28)	&			-0.087			&			0.033 			&			0.13      				&			4.83    			&		0.76      				\\
	5,-25   &		-0.4196  (0.02)				&		6.081			(0.09)	&			3.883		(0.18)	&			-0.102			&			0.034 			&			0.14      				&			5.04    			&		0.83      				\\
	10,-25   &		-0.5856  (0.02)				&		6.525			(0.07)	&			3.599		(0.16)	&			-0.153			&			0.043 			&			0.16      				&			5.32    			&		1.09      				\\
	15,-25   &		-0.6487  (0.02)				&		6.611			(0.06)	&			3.69 		(0.14)	&			-0.165			&			0.056 			&			0.13      				&			4.48    			&		1.42      				\\
	20,-25   &		-0.4641  (0.02)				&		6.592			(0.07)	&			3.239		(0.17)	&			-0.135			&			0.055 			&			0.14      				&			4.19    			&		1.11      				\\
	25,-25   &		-0.387   (0.02)				&		6.872			(0.07)	&			2.765		(0.16)	&			-0.131			&			0.058 			&			0.16      				&			4.09    			&		0.96      				\\
	30,-25   &		-0.4065  (0.02)				&		6.844			(0.06)	&			2.742		(0.13)	&			-0.139			&			0.067 			&			0.16      				&			4.22    			&		0.99      				\\
	35,-25   &		-0.2898  (0.02)				&		6.742			(0.07)	&			2.485		(0.15)	&			-0.110			&			0.072 			&			0.19      				&			4.52    			&		0.7       				\\
	-15,-20   &		-0.2043  (0.02)				&		6.119			(0.24)	&			4.444		(0.49)	&			-0.043			&			-      			&			0.06      				&			2.44    			&		0.76      				\\
	-10,-20   &		-0.2911  (0.02)				&		6.787			(0.17)	&			4.375		(0.38)	&			-0.063			&			0.032 			&			0.1       				&			4.12    			&		0.69      				\\
	-5,-20   &		-0.3486  (0.03)				&		6.587			(0.16)	&			4.421		(0.39)	&			-0.074			&			0.038 			&			0.09      				&			3.91    			&		0.86      				\\
	0,-20   &		-0.3905  (0.03)				&		6.243			(0.15)	&			4.351		(0.31)	&			-0.084			&			0.033 			&			0.11      				&			4.42    			&		0.86      				\\
	5,-20   &		-0.5314  (0.03)				&		6.276			(0.12)	&			4.039		(0.31)	&			-0.124			&			0.034 			&			0.16      				&			6.17    			&		0.86      				\\
	10,-20   &		-0.6238  (0.02)				&		6.413			(0.07)	&			3.663		(0.17)	&			-0.160			&			0.052 			&			0.18      				&			6.35    			&		1         				\\
	15,-20   &		-0.5478  (0.02)				&		6.644			(0.04)	&			3.077		(0.10)	&			-0.167			&			0.076 			&			0.17      				&			4.82    			&		1.16      				\\
	20,-20   &		-0.4843  (0.02)				&		6.485			(0.06)	&			2.919		(0.16)	&			-0.156			&			0.091 			&			0.16      				&			4.37    			&		1.15      				\\
	25,-20   &		-0.416   (0.02)				&		6.98 			(0.07)	&			2.798		(0.15)	&			-0.140			&			0.088 			&			0.16      				&			4.22    			&		1.03      				\\
	30,-20   &		-0.3507  (0.02)				&		6.755			(0.07)	&			2.573		(0.17)	&			-0.128			&			0.085 			&			0.18      				&			4.29    			&		0.87      				\\
	35,-20   &		-0.2327  (0.02)				&		7.052			(0.13)	&			2.698		(0.36)	&			-0.081			&			0.077 			&			0.16      				&			4.06    			&		0.63      				\\
	40,-20   &		-0.1936  (0.02)				&		7.092			(0.16)	&			2.619		(0.40)	&			-0.069			&			0.072 			&			0.13      				&			3.27    			&		0.63      				\\
	-15,-15   &		-0.18    (0.02)				&		5.807			(0.37)	&			5.28 		(0.79)	&			-0.032			&			0.021 			&			0.07      				&			3.34    			&		0.51      				\\
	-10,-15   &		-0.2576  (0.02)				&		5.85 			(0.13)	&			3.681		(0.33)	&			-0.066			&			0.034 			&			0.12      				&			4.08    			&		0.62      				\\
	-5,-15   &		-0.3863  (0.02)				&		5.914			(0.08)	&			3.561		(0.20)	&			-0.102			&			0.039 			&			0.13      				&			4.39    			&		0.87      				\\
	0,-15   &		-0.5037  (0.03)				&		6.333			(0.10)	&			4.047		(0.26)	&			-0.117			&			0.031 			&			0.14      				&			5.26    			&		0.94      				\\
	5,-15   &		-0.6316  (0.02)				&		6.156			(0.06)	&			3.962		(0.17)	&			-0.150			&			0.041 			&			0.18      				&			6.72    			&		0.95      				\\
	10,-15   &		-0.6291  (0.02)				&		6.35 			(0.05)	&			3.592		(0.13)	&			-0.165			&			0.070 			&			0.21      				&			7.21    			&		0.93      				\\
	15,-15   &		-0.4989  (0.01)				&		6.824			(0.04)	&			2.981		(0.11)	&			-0.157			&			0.125 			&			0.19      				&			5.25    			&		1.04      				\\
	20,-15   &		-0.4105  (0.02)				&		6.802			(0.07)	&			2.693		(0.16)	&			-0.143			&			0.189 			&			0.25      				&			6.43    			&		0.83      				\\
	25,-15   &		-0.3963  (0.02)				&		6.883			(0.06)	&			2.811		(0.15)	&			-0.132			&			0.168 			&			0.19      				&			4.98    			&		0.94      				\\
	30,-15   &		-0.3329  (0.02)				&		7.104			(0.07)	&			2.96 		(0.19)	&			-0.106			&			0.135 			&			0.15      				&			4.15    			&		0.9       				\\
	40,-15   &		-0.1788  (0.02)				&		7.287			(0.21)	&			3.214		(0.58)	&			-0.052			&			0.081 			&			0.1       				&			3.16    			&		0.61      				\\
	-15,-10   &		-0.2646  (0.02)				&		5.313			(0.12)	&			3.873		(0.33)	&			-0.064			&			0.026 			&			0.19      				&			6.81    			&		0.4       				\\
	-10,-10   &		-0.3038  (0.02)				&		5.827			(0.11)	&			3.162		(0.25)	&			-0.090			&			0.035 			&			0.24      				&			7.09    			&		0.46      				\\
	-5,-10   &		-0.5445  (0.03)				&		5.982			(0.08)	&			3.54 		(0.19)	&			-0.144			&			0.034 			&			0.18      				&			5.83    			&		0.93      				\\
	0,-10   &		-0.6587  (0.02)				&		6.055			(0.06)	&			3.713		(0.16)	&			-0.167			&			0.036 			&			0.19      				&			6.61    			&		1         				\\
	5,-10   &		-0.6776  (0.02)				&		6.162			(0.06)	&			3.597		(0.15)	&			-0.177			&			0.048 			&			0.2       				&			6.79    			&		1.02      				\\
	10,-10   &		-0.6345  (0.03)				&		6.208			(0.06)	&			3.215		(0.15)	&			-0.185			&			0.105 			&			0.22      				&			6.80    			&		1.02      				\\
	15,-10   &		-0.5845  (0.02)				&		6.538			(0.04)	&			3.125		(0.12)	&			-0.176			&			0.449 			&			0.2       				&			5.84    			&		1.43      				\\
	20,-10   &		-0.6362  (0.03)				&		7.295			(0.06)	&			3.039		(0.15)	&			-0.197			&			1.257 			&			0.27      				&			7.59    			&		2.1       				\\
	25,-10   &		-0.5281  (0.03)				&		7.238			(0.07)	&			2.84 		(0.17)	&			-0.175			&			1.093 			&			0.15      				&			4.06    			&		2.33      				\\
	30,-10   &		-0.3295  (0.02)				&		7.465			(0.08)	&			2.657		(0.20)	&			-0.116			&			0.314 			&			0.16      				&			3.89    			&		1.12      				\\
	35,-10   &		-0.2776  (0.02)				&		7.476			(0.10)	&			2.296		(0.21)	&			-0.114			&			0.148 			&			0.19      				&			4.01    			&		0.82      				\\
	-15,-5    &		-0.3798  (0.02)				&		5.355			(0.09)	&			4.264		(0.27)	&			-0.084			&			0.030 			&			0.22      				&			8.69    			&		0.46      				\\
	-10,-5    &		-0.45    (0.02)				&		5.968			(0.07)	&			3.508		(0.18)	&			-0.121			&			0.038 			&			0.22      				&			7.14    			&		0.66      				\\
	-5,-5    &		-0.6701  (0.02)				&		6.133			(0.06)	&			3.641		(0.16)	&			-0.173			&			0.038 			&			0.18      				&			6.21    			&		1.08      				\\
	0,-5    &		-0.8745  (0.02)				&		6.01 			(0.04)	&			3.509		(0.10)	&			-0.234			&			0.042 			&			0.22      				&			7.34    			&		1.21      				\\
	5,-5    &		-0.8472  (0.02)				&		6.016			(0.04)	&			3.376		(0.10)	&			-0.236			&			0.065 			&			0.23      				&			7.32    			&		1.21      				\\
	10,-5    &		-0.7178  (0.02)				&		6.341			(0.05)	&			3.637		(0.13)	&			-0.185			&			0.285 			&			0.21      				&			7.03    			&		1.28      				\\
	15,-5    &		-0.7391  (0.03)				&		6.608			(0.06)	&			3.475		(0.15)	&			-0.200			&			1.645 			&			0.37      				&			12.10   			&		2.29      				\\
	20,-5    &		-1.097   (0.03)				&		7.125			(0.05)	&			3.275		(0.13)	&			-0.315			&			3.327 			&			0.29      				&			8.79    			&		4.59      				\\
	25,-5    &		-0.9857  (0.02)				&		7.261			(0.04)	&			2.95 		(0.09)	&			-0.314			&			2.591 			&			0.15      				&			4.24    			&		4.8       				\\
	30,-5    &		-0.4751  (0.02)				&		7.251			(0.06)	&			2.532		(0.15)	&			-0.176			&			0.594 			&			0.13      				&			2.99    			&		2.08      				\\
	35,-5    &		-0.2984  (0.02)				&		7.612			(0.09)	&			2.313		(0.19)	&			-0.121			&			0.175 			&			0.18      				&			3.97    			&		0.9       				\\
	-20,0     &		-0.271   (0.02)				&		5.339			(0.14)	&			4.025		(0.32)	&			-0.063			&			0.018 			&			0.13      				&			5.09    			&		0.52      				\\
	-15,0     &		-0.3863  (0.02)				&		5.753			(0.10)	&			4.431		(0.25)	&			-0.082			&			0.031 			&			0.2       				&			8.43    			&		0.48      				\\
	-10,0     &		-0.5059  (0.02)				&		5.647			(0.08)	&			3.831		(0.19)	&			-0.124			&			0.039 			&			0.22      				&			7.90    			&		0.67      				\\
	-5,0     &		-0.8185  (0.02)				&		6.122			(0.05)	&			3.973		(0.12)	&			-0.194			&			0.041 			&			0.19      				&			7.22    			&		1.14      				\\
	0,0     &		-1.007   (0.02)				&		6.216			(0.03)	&			3.392		(0.07)	&			-0.279			&			0.045 			&			0.26      				&			8.21    			&		1.27      				\\
	5,0     &		-0.9486  (0.02)				&		6.4  			(0.03)	&			3.299		(0.08)	&			-0.270			&			0.090 			&			0.21      				&			6.59    			&		1.5       				\\
	10,0     &		-0.86    (0.03)				&		6.425			(0.06)	&			3.757		(0.16)	&			-0.215			&			0.391 			&			0.18      				&			6.43    			&		1.68      				\\
	15,0     &		-0.7699  (0.02)				&		6.679			(0.04)	&			3.431		(0.10)	&			-0.211			&			1.617 			&			0.47      				&			15.10   			&		2.18      				\\
	20,0     &		-1.089   (0.03)				&		7.051			(0.04)	&			3.188		(0.11)	&			-0.321			&			3.179 			&			0.35      				&			10.40   			&		4.27      				\\
	25,0     &		-1.027   (0.02)				&		7.212			(0.03)	&			2.994		(0.08)	&			-0.322			&			2.477 			&			0.38      				&			10.70   			&		3.5       				\\
	30,0     &		-0.5105  (0.02)				&		7.433			(0.05)	&			2.619		(0.13)	&			-0.183			&			0.593 			&			0.29      				&			7.24    			&		1.31      				\\
	35,0     &		-0.3393  (0.02)				&		7.531			(0.09)	&			2.574		(0.20)	&			-0.124			&			0.134 			&			0.17      				&			4.21    			&		0.91      				\\
	-25,5     &		-0.2438  (0.02)				&		5.342			(0.13)	&			3.16 		(0.27)	&			-0.072			&			0.013 			&			0.14      				&			4.03    			&		0.58      				\\
	-20,5     &		-0.3443  (0.02)				&		6.028			(0.11)	&			3.645		(0.28)	&			-0.089			&			0.019 			&			0.15      				&			5.05    			&		0.67      				\\
	-15,5     &		-0.408   (0.01)				&		6.167			(0.06)	&			3.627		(0.14)	&			-0.106			&			0.028 			&			0.21      				&			7.15    			&		0.59      				\\
	-10,5     &		-0.5894  (0.02)				&		5.994			(0.07)	&			3.658		(0.17)	&			-0.151			&			0.039 			&			0.24      				&			8.29    			&		0.74      				\\
	-5,5     &		-0.8586  (0.03)				&		6.065			(0.05)	&			3.27 		(0.12)	&			-0.247			&			0.041 			&			0.27      				&			8.18    			&		1.1       				\\
	0,5     &		-1.063   (0.02)				&		6.245			(0.03)	&			3.331		(0.07)	&			-0.300			&			0.048 			&			0.28      				&			8.83    			&		1.27      				\\
	5,5     &		-1.102   (0.02)				&		6.414			(0.03)	&			3.341		(0.08)	&			-0.310			&			0.076 			&			0.26      				&			8.12    			&		1.44      				\\
	10,5     &		-1.041   (0.03)				&		6.566			(0.05)	&			3.551		(0.11)	&			-0.276			&			0.180 			&			0.23      				&			7.69    			&		1.52      				\\
	15,5     &		-0.8165  (0.02)				&		6.92 			(0.05)	&			3.45 		(0.11)	&			-0.222			&			0.583 			&			0.32      				&			10.50   			&		1.39      				\\
	20,5     &		-0.7599  (0.02)				&		7.199			(0.05)	&			3.208		(0.12)	&			-0.223			&			1.075 			&			0.62      				&			18.70   			&		1.56      				\\
	25,5     &		-0.7333  (0.02)				&		7.22 			(0.04)	&			3.101		(0.10)	&			-0.222			&			0.822 			&			0.42      				&			12.30   			&		1.47      				\\
	30,5     &		-0.6069  (0.02)				&		7.577			(0.04)	&			2.89 		(0.11)	&			-0.197			&			0.245 			&			0.28      				&			7.50    			&		1.06      				\\
	35,5     &		-0.3872  (0.02)				&		7.632			(0.08)	&			2.412		(0.18)	&			-0.151			&			0.110 			&			0.19      				&			4.36    			&		0.97      				\\
	-30,10    &		-0.1847  (0.02)				&		5.655			(0.15)	&			2.883		(0.40)	&			-0.060			&			0.005 			&			0.11      				&			2.97    			&		0.59      				\\
	-25,10    &		-0.3499  (0.02)				&		5.69 			(0.09)	&			3.304		(0.22)	&			-0.100			&			0.013 			&			0.15      				&			4.78    			&		0.71      				\\
	-20,10    &		-0.551   (0.03)				&		5.757			(0.09)	&			4.103		(0.28)	&			-0.126			&			0.017 			&			0.15      				&			5.60    			&		0.95      				\\
	-15,10    &		-0.6083  (0.02)				&		5.947			(0.08)	&			4.071		(0.21)	&			-0.140			&			0.026 			&			0.23      				&			8.66    			&		0.72      				\\
	-10,10    &		-0.6509  (0.02)				&		5.769			(0.05)	&			3.274		(0.12)	&			-0.187			&			0.038 			&			0.31      				&			9.61    			&		0.74      				\\
	-5,10    &		-0.879   (0.02)				&		6.083			(0.04)	&			3.276		(0.11)	&			-0.252			&			0.041 			&			0.28      				&			8.62    			&		1.07      				\\
	0,10    &		-0.9472  (0.02)				&		6.276			(0.04)	&			3.373		(0.09)	&			-0.264			&			0.037 			&			0.26      				&			8.10    			&		1.21      				\\
	5,10    &		-0.952   (0.02)				&		6.468			(0.04)	&			3.408		(0.11)	&			-0.262			&			0.050 			&			0.27      				&			8.52    			&		1.17      				\\
	10,10    &		-0.9794  (0.03)				&		6.587			(0.04)	&			3.569		(0.11)	&			-0.258			&			0.085 			&			0.25      				&			8.47    			&		1.24      				\\
	15,10    &		-0.6987  (0.02)				&		7.048			(0.04)	&			3.044		(0.09)	&			-0.216			&			0.122 			&			0.32      				&			9.19    			&		0.91      				\\
	20,10    &		-0.5961  (0.02)				&		7.577			(0.05)	&			2.802		(0.12)	&			-0.200			&			0.175 			&			0.36      				&			9.56    			&		0.83      				\\
	25,10    &		-0.5113  (0.02)				&		7.537			(0.05)	&			2.569		(0.12)	&			-0.187			&			0.163 			&			0.36      				&			8.81    			&		0.77      				\\
	30,10    &		-0.6237  (0.03)				&		7.788			(0.05)	&			2.662		(0.14)	&			-0.220			&			0.136 			&			0.24      				&			6.08    			&		1.16      				\\
	35,10    &		-0.4458  (0.02)				&		8.229			(0.05)	&			2.416		(0.12)	&			-0.173			&			0.086 			&			0.21      				&			4.86    			&		0.98      				\\
	-30,15    &		-7.97E-02(0.01)				&		5.556			(0.12)	&			1.537		(0.31)	&			-0.049			&			0.005 			&			0.14      				&			1.96    			&		0.39      				\\
	-25,15    &		-0.1822  (0.02)				&		5.638			(0.10)	&			2.622		(0.31)	&			-0.065			&			0.012 			&			0.12      				&			3.02    			&		0.58      				\\
	-20,15    &		-0.284   (0.01)				&		6.01 			(0.07)	&			2.793		(0.16)	&			-0.096			&			0.017 			&			0.18      				&			4.72    			&		0.6       				\\
	-15,15    &		-0.5231  (0.01)				&		5.877			(0.04)	&			3.2  		(0.10)	&			-0.154			&			0.022 			&			0.31      				&			9.41    			&		0.6       				\\
	-10,15    &		-0.6096  (0.01)				&		5.868			(0.03)	&			3.21 		(0.07)	&			-0.178			&			0.031 			&			0.3       				&			9.03    			&		0.72      				\\
	-5,15    &		-0.737   (0.02)				&		6.181			(0.03)	&			3.195		(0.08)	&			-0.217			&			0.036 			&			0.29      				&			8.69    			&		0.9       				\\
	0,15    &		-0.7129  (0.02)				&		6.411			(0.04)	&			3.166		(0.08)	&			-0.212			&			0.033 			&			0.26      				&			7.68    			&		0.96      				\\
	5,15    &		-0.6118  (0.02)				&		6.528			(0.05)	&			3.285		(0.12)	&			-0.175			&			0.034 			&			0.2       				&			6.08    			&		1.01      				\\
	10,15    &		-0.5221  (0.01)				&		6.819			(0.04)	&			3.136		(0.10)	&			-0.156			&			0.043 			&			0.18      				&			5.22    			&		1         				\\
	15,15    &		-0.4577  (0.02)				&		6.879			(0.05)	&			3.173		(0.13)	&			-0.136			&			0.059 			&			0.25      				&			7.36    			&		0.68      				\\
	20,15    &		-0.4603  (0.02)				&		7.154			(0.07)	&			3.133		(0.18)	&			-0.138			&			0.077 			&			0.26      				&			7.73    			&		0.68      				\\
	25,15    &		-0.5063  (0.01)				&		7.285			(0.04)	&			3.016		(0.11)	&			-0.158			&			0.077 			&			0.3       				&			8.37    			&		0.7       				\\
	30,15    &		-0.5156  (0.02)				&		7.391			(0.05)	&			3.025		(0.12)	&			-0.160			&			0.061 			&			0.26      				&			7.29    			&		0.77      				\\
	35,15    &		-0.473   (0.02)				&		7.632			(0.06)	&			3.264		(0.15)	&			-0.136			&			0.048 			&			0.16      				&			5.02    			&		0.95      				\\
	40,15    &		-0.3587  (0.02)				&		8.072			(0.08)	&			2.938		(0.24)	&			-0.115			&			0.046 			&			0.16      				&			4.45    			&		0.82      				\\
	-25,20    &		-8.56E-02(0.01)				&		6.378			(0.13)	&			1.792		(0.27)	&			-0.045			&			0.008 			&			0.21      				&			3.58    			&		0.24      				\\
	-20,20    &		-0.2295  (0.02)				&		6.244			(0.08)	&			2.34 		(0.18)	&			-0.092			&			0.014 			&			0.25      				&			5.50    			&		0.43      				\\
	-15,20    &		-0.4757  (0.01)				&		6.112			(0.04)	&			3.012		(0.10)	&			-0.148			&			0.019 			&			0.37      				&			10.60   			&		0.49      				\\
	-10,20    &		-0.5878  (0.01)				&		6.006			(0.04)	&			2.839		(0.08)	&			-0.195			&			0.024 			&			0.34      				&			9.13    			&		0.7       				\\
	-5,20    &		-0.4776  (0.02)				&		6.255			(0.06)	&			2.81 		(0.15)	&			-0.160			&			0.027 			&			0.26      				&			6.76    			&		0.74      				\\
	0,20    &		-0.3976  (0.02)				&		6.323			(0.06)	&			2.697		(0.13)	&			-0.138			&			0.030 			&			0.22      				&			5.70    			&		0.72      				\\
	5,20    &		-0.2805  (0.02)				&		6.471			(0.08)	&			2.734		(0.19)	&			-0.096			&			0.033 			&			0.23      				&			5.82    			&		0.51      				\\
	10,20    &		-0.2812  (0.02)				&		6.831			(0.08)	&			2.741		(0.18)	&			-0.096			&			0.036 			&			0.2       				&			5.03    			&		0.58      				\\
	15,20    &		-0.2583  (0.01)				&		6.698			(0.09)	&			2.994		(0.20)	&			-0.081			&			0.040 			&			0.18      				&			5.16    			&		0.52      				\\
	20,20    &		-0.2941  (0.03)				&		7.105			(0.14)	&			3.265		(0.38)	&			-0.085			&			0.035 			&			0.2       				&			6.08    			&		0.51      				\\
	25,20    &		-0.3525  (0.01)				&		6.743			(0.06)	&			2.994		(0.14)	&			-0.111			&			0.033 			&			0.23      				&			6.61    			&		0.56      				\\
	30,20    &		-0.4116  (0.02)				&		7.142			(0.06)	&			3.144		(0.14)	&			-0.123			&			0.029 			&			0.29      				&			8.67    			&		0.51      				\\
	35,20    &		-0.3966  (0.02)				&		7.617			(0.10)	&			3.869		(0.28)	&			-0.096			&			0.028 			&			0.16      				&			5.73    			&		0.69      				\\
	-25,25    &		-0.119   (0.02)				&		6.228			(0.14)	&			2.038		(0.34)	&			-0.055			&			0.007 			&			0.27      				&			5.21    			&		0.24      				\\
	-20,25    &		-0.1881  (0.02)				&		6.228			(0.10)	&			2.128		(0.20)	&			-0.083			&			0.012 			&			0.31      				&			6.26    			&		0.32      				\\
	-15,25    &		-0.4305  (0.02)				&		6.056			(0.05)	&			3.147		(0.13)	&			-0.129			&			0.017 			&			0.26      				&			7.81    			&		0.57      				\\
	-10,25    &		-0.4385  (0.02)				&		5.88 			(0.05)	&			2.623		(0.13)	&			-0.157			&			0.022 			&			0.27      				&			6.73    			&		0.68      				\\
	-5,25    &		-0.346   (0.02)				&		6.456			(0.09)	&			2.854		(0.19)	&			-0.114			&			0.023 			&			0.19      				&			5.17    			&		0.67      				\\
	0,25    &		-0.2902  (0.02)				&		6.753			(0.11)	&			3.032		(0.25)	&			-0.090			&			0.028 			&			0.18      				&			5.04    			&		0.58      				\\
	5,25    &		-0.1657  (0.02)				&		6.667			(0.15)	&			2.502		(0.28)	&			-0.062			&			0.031 			&			0.2       				&			4.67    			&		0.38      				\\
	10,25    &		-0.1609  (0.02)				&		6.448			(0.12)	&			2.629		(0.29)	&			-0.058			&			0.033 			&			0.16      				&			3.91    			&		0.43      				\\
	15,25    &		-0.1681  (0.02)				&		6.763			(0.14)	&			2.926		(0.31)	&			-0.054			&			0.034 			&			0.14      				&			3.80    			&		0.45      				\\
	25,25    &		-0.283   (0.02)				&		6.627			(0.12)	&			3.526		(0.30)	&			-0.075			&			0.031 			&			0.15      				&			5.12    			&		0.56      				\\
	30,25    &		-0.3594  (0.03)				&		7.174			(0.14)	&			3.741		(0.35)	&			-0.090			&			0.029 			&			0.19      				&			6.67    			&		0.55      				\\
	-30,30    &		-6.26E-02(0.01)				&		5.851			(0.12)	&			1.312		(0.26)	&			-0.045			&			 -     			&			0.29      				&			3.53    			&		0.18      				\\
	-25,30    &		-0.1724  (0.03)				&		6.01 			(0.24)	&			3.71 		(0.80)	&			-0.044			&			0.004 			&			0.13      				&			4.58    			&		0.36      				\\
	-20,30    &		-0.1762  (0.02)				&		6.304			(0.08)	&			1.821		(0.18)	&			-0.091			&			0.010 			&			0.31      				&			5.39    			&		0.35      				\\
	-15,30    &		-0.3506  (0.02)				&		6.004			(0.07)	&			2.822		(0.16)	&			-0.117			&			0.015 			&			0.22      				&			5.86    			&		0.61      				\\
	-10,30    &		-0.3671  (0.02)				&		5.999			(0.07)	&			2.835		(0.15)	&			-0.122			&			0.016 			&			0.22      				&			5.90    			&		0.63      				\\
	-5,30    &		-0.2804  (0.02)				&		6.453			(0.10)	&			2.913		(0.18)	&			-0.090			&			0.017 			&			0.2       				&			5.54    			&		0.51      				\\
	0,30    &		-0.2279  (0.02)				&		6.54 			(0.14)	&			3.179		(0.28)	&			-0.067			&			0.020 			&			0.18      				&			5.48    			&		0.42      				\\
	5,30    &		-0.1997  (0.02)				&		6.763			(0.16)	&			2.979		(0.32)	&			-0.063			&			0.024 			&			0.21      				&			5.91    			&		0.35      				\\
	15,30    &		-0.1365  (0.02)				&		6.272			(0.19)	&			3.002		(0.41)	&			-0.043			&			0.027 			&			0.12      				&			3.43    			&		0.4       				\\
	20,30    &		-0.1653  (0.02)				&		6.296			(0.21)	&			3.424		(0.45)	&			-0.045			&			0.033 			&			0.1       				&			3.27    			&		0.5       				\\
	25,30    &		-0.2251  (0.02)				&		7.235			(0.14)	&			2.851		(0.32)	&			-0.074			&			0.036 			&			0.15      				&			3.96    			&		0.58      				\\
	30,30    &		-0.2934  (0.03)				&		7.528			(0.20)	&			4.416		(0.42)	&			-0.062			&			0.032 			&			0.15      				&			6.15    			&		0.49      				\\
	35,30    &		-0.3368  (0.03)				&		8.223			(0.21)	&			4.904		(0.69)	&			-0.065			&			0.018 			&			0.14      				&			6.47    			&		0.51      				\\
	40,30    &		-0.3056  (0.02)				&		8.698			(0.12)	&			3.06 		(0.30)	&			-0.094			&			0.011 			&			0.25      				&			7.23    			&		0.43      				\\
	-20,35    &		-0.122   (0.01)				&		5.89 			(0.11)	&			2.093		(0.23)	&			-0.055			&			0.006 			&			0.24      				&			4.66    			&		0.27      				\\
	-15,35    &		-0.3171  (0.02)				&		6.122			(0.09)	&			2.985		(0.25)	&			-0.100			&			0.009 			&			0.2       				&			5.66    			&		0.56      				\\
	-10,35    &		-0.3227  (0.01)				&		5.761			(0.06)	&			2.538		(0.13)	&			-0.119			&			0.011 			&			0.25      				&			5.88    			&		0.56      				\\
	-5,35    &		-0.2316  (0.02)				&		6.217			(0.13)	&			3.267		(0.25)	&			-0.067			&			0.013 			&			0.16      				&			5.02    			&		0.45      				\\
	5,35    &		-0.2155  (0.02)				&		6.965			(0.14)	&			3.321		(0.33)	&			-0.061			&			0.016 			&			0.16      				&			5.06    			&		0.42      				\\
	15,35    &		-0.1827  (0.02)				&		6.697			(0.17)	&			3.237		(0.42)	&			-0.053			&			0.017 			&			0.12      				&			3.65    			&		0.49      				\\
	20,35    &		-0.3746  (0.05)				&		6.711			(0.53)	&			9.407		(1.95)	&			-0.037			&			 -     			&			0.04      				&			3.10    			&		1.08      				\\
	25,35    &		-0.2111  (0.02)				&		7.523			(0.18)	&			3.495		(0.43)	&			-0.057			&			0.022 			&			0.12      				&			4.03    			&		0.51      				\\
	30,35    &		-0.2988  (0.02)				&		7.231			(0.15)	&			3.975		(0.32)	&			-0.071			&			0.019 			&			0.15      				&			5.56    			&		0.53      				\\
	35,35    &		-0.4561  (0.02)				&		7.969			(0.09)	&			3.785		(0.25)	&			-0.113			&			0.011 			&			0.19      				&			6.74    			&		0.67      				\\
	40,35    &		-0.4512  (0.03)				&		8.458			(0.11)	&			3.865		(0.29)	&			-0.110			&			0.010 			&			0.23      				&			8.49    			&		0.54      				\\
	-15,40    &		-0.1673  (0.02)				&		6.02 			(0.12)	&			2.456		(0.30)	&			-0.064			&			0.004 			&			0.16      				&			3.79    			&		0.43      				\\
	-10,40    &		-0.241   (0.01)				&		5.977			(0.08)	&			2.744		(0.18)	&			-0.083			&			0.007 			&			0.22      				&			5.59    			&		0.43      				\\
	5,40    &		-0.1615  (0.02)				&		6.872			(0.19)	&			3.201		(0.45)	&			-0.047			&			0.012 			&			0.15      				&			4.46    			&		0.36      				\\
	10,40    &		-0.1639  (0.02)				&		7.148			(0.15)	&			2.775		(0.31)	&			-0.056			&			0.011 			&			0.16      				&			4.14    			&		0.39      				\\
	15,40    &		-0.2213  (0.01)				&		7.189			(0.11)	&			3.474		(0.26)	&			-0.060			&			0.012 			&			0.11      				&			3.73    			&		0.57      				\\
	20,40    &		-0.3022  (0.02)				&		6.847			(0.16)	&			4.046		(0.40)	&			-0.070			&			0.014 			&			0.09      				&			3.45    			&		0.82      				\\
	25,40    &		-0.2829  (0.02)				&		7.331			(0.12)	&			3.435		(0.29)	&			-0.077			&			0.013 			&			0.13      				&			4.10    			&		0.66      				\\
	30,40    &		-0.3546  (0.02)				&		7.693			(0.07)	&			3.17 		(0.17)	&			-0.105			&			0.011 			&			0.19      				&			5.71    			&		0.61      				\\
	35,40    &		-0.4666  (0.03)				&		8.02 			(0.09)	&			3.417		(0.26)	&			-0.128			&			0.011 			&			0.22      				&			7.06    			&		0.66      				\\
	40,40    &		-0.3529  (0.02)				&		7.929			(0.09)	&			3.038		(0.25)	&			-0.109			&			0.012 			&			0.2       				&			5.76    			&		0.61      				\\
	\enddata
	\tablecomments{The central position is 18:30:03 -2:02:40 (EQ J2000).}
\end{deluxetable}

\end{document}